\documentclass[iop]{emulateapj}
\usepackage{graphicx,fleqn,times,color}
\usepackage{enumitem}

\def\aj{AJ}
\def\apj{ApJ}

\def\mnras{MNRAS}

\def\today{\ifcase\month\or
 January\or February\or March\or April\or May\or June\or
 July\or August\or September\or October\or November\or
 December\fi\space\number\day, \number\year}

\def\todmy{\number\day\space\ifcase\month\or
 January\or February\or March\or April\or May\or June\or
 July\or August\or September\or October\or November\or
 December\fi\space\number\year}

\received{}
\revised{}
\accepted{}

\begin{document}

\title{Forming disk
  galaxies in wet major mergers. I. Three fiducial examples 
  }
\author{E.~Athanassoula, S.~A.~Rodionov,
  N. Peschken, J. C. Lambert}
\affil{Laboratoire d'Astrophysique de Marseille (LAM), UMR7326,
CNRS/Aix Marseille Universit\'e, Technop\^ole de Marseille-Etoile, \\
38 rue Fr\'ed\'eric Joliot Curie, 13388 Marseille C\'edex 13,
France\\}
\email{lia@lam.fr}

\shorttitle{Forming disk
  galaxies in wet major mergers I Three fiducial examples with
  bars.}
\shortauthors{Athanassoula et al.}

\begin{abstract}
Using three fiducial Nbody+SPH simulations, we
follow the merging of two disk galaxies with a
hot gaseous halo component each, and examine whether the
merger remnant can be a spiral galaxy. The stellar progenitor disks are
destroyed by violent relaxation during the merging and most of their stars
form a classical bulge, while the remaining form a
thick disk and its bar.  
A new stellar disk forms subsequently and gradually in the remnant
from the gas accreted mainly from the halo. It is vertically thin and well
extended in its equatorial plane. A bar starts forming before the disk
is fully in place, contrary to what is assumed in idealised
simulations of isolated bar-forming galaxies. It has 
morphological features such as ansae and boxy/peanut
bulges. Stars of different ages populate 
different parts of the box/peanut. 
A disky pseudobulge forms also, so that by the end of the
simulation, all three types of bulges coexist.
The oldest stars are found in the classical bulge, followed by those
of the thick disk, then by those in the thin disk. The
youngest stars are in the spiral arms and the disky pseudobulge.
The disk surface
density profiles are of type II (exponential with downbending),
and the circular velocity curves are flat and show that the disks are submaximum in
these examples: two clearly so and one near-borderline between maximum
and submaximum. On 
average, only roughly between 10 and 20\% of the stellar mass is 
in the classical bulge of the final models, i.e. much less than in previous
simulations. \end{abstract}


\section{Introduction}
\label{sec:intro}
\indent

What results from a merger of two disk galaxies of comparable mass?
\citet{Toomre.Toomre.72} were the first to propose that such a merger 
will form an elliptical. After some strong initial rebuttals, this
was generally accepted (see e.g. \citealt{Barnes.98} and \citealt{Schweizer.98}
for reviews), 
only to be questioned again in the last decade. 
Indeed, several observations at intermediate
redshifts, suggest that the result of a merging of two gas-rich disk
galaxies of comparable luminosity is actually not an elliptical, but a disk
galaxy \citep[e.g.][]{Hammer.FEZLC.05, Hammer.FPYARD.09,
  Hammer.FYAPRP.09}. Roughly concurrently, pioneering and seminal
numerical simulations showed that the remnants of mergers of gas-rich
disk galaxies do have a disk component
(e.g. \citealt{Barnes.02}, \citealt{Springel.Hernquist.05}, 
\citealt{Cox.JPS.06}, \citealt{Robertson.BCDMHSY.06a}, 
\citealt{Lotz.JCP.10b},
\citealt{Governato.BBMWJSPCWQ.09}, \citealt{Hopkins.CYH.09,  
  Hopkins.CHNHM.13}, \citealt{Wang.HAPYF.12}, \citealt{Borlaff.EMRPQTPGZGB.14},
\citealt{Querejeta.EMTBRPZG.15}).  
However, the relative mass and/or extent of these disks are in general
considerably smaller than those of present day 
spiral galaxy disks. This formation mechanism could thus perhaps be
appropriate for 
lenticulars, but not for spirals. Furthermore, although bars are
present in about two thirds of disk galaxies in the local universe
\citep[][etc.]{Eskridge.P.00, Knapen.SP.00,
  Menendez.DSSJS.07, Buta.P.15} and are believed to be a
major driver of secular evolution \citep[see
  e.g.][for reviews of the theoretical and observational parts,
  respectively]{Athanassoula.13, Kormendy.13}, these works provide 
little or no
information on whether and when major merger remnants can have bars and on
whether their bar properties and parameters are realistic.

It is thus necessary to return to this still open question and to test
whether the remnants of major mergers could be spiral galaxies with
the appropriate 
morphology, mass distribution, kinematics and substructures. In this paper, the first
of a series, 
we present first results for three fiducial wet major merger simulations
and their remnants. We first present the improvements we introduce
in the modeling of the protogalaxies and in the simulation resolution
and give information on the code we used and
  our initial conditions
  (Sect.~\ref{sec:sims}). Sect.~\ref{sec:results} presents and
  discusses our results. We show that the disk of
the remnant can extend over several scalelengths, while the $B/T$ ratio 
(classical-bulge-to-total stellar mass ratio) can reach sufficiently
low values to be representative of spirals.  We also
discuss morphologies, including those of spirals and bars, show rotation
curves and projected density radial profiles, and discuss when and how
the various 
components are produced. We compare the morphology and
kinematics of stellar populations with different ages
(Sect.~\ref{subsec:agesplit_res}) and also
present a simple scenario for the formation of disk galaxies
(Sect.~\ref{sec:scenario}), before summarizing and concluding
(Sect.~\ref{sec:summary}). The nomenclature used in this paper is
summarized in the Appendix. For brevity, we will hereafter 
often refer to our simulated galaxies simply as ``galaxies'', and to
stellar particles in the simulation as ``stars''.


\section{Simulations}
\label{sec:sims}

\subsection{General context}
\label{subsec:general}

Compared to previous simulations on this specific subject, ours present
one numerical and one conceptual advantage.

As a numerical advantage, we have a much better resolution than
previous simulations tackling 
the specific question that we have set out to address in this project,
namely whether major mergers of gas rich disk galaxies can form spirals, or
not. Indeed, to answer this question necessitates a survey of simulations, as 
e.g. in \cite{Hopkins.CYH.09}, so that the use of high
resolution becomes computationally much more expensive than for problems
necessitating only  
few simulations. For this reason previous attempts restricted themselves to
considerably less than or of the order of half a million particles in
total.\footnote{Amongst the simulations with a resolution yet higher
  than ours we note two simulations of Milky Way sized galaxies with a
resolution of four pc, i.e. roughly six times better than ours
\citep{Hopkins.CHNHM.13}.  
Such a resolution, however, would be impossible for our project, since
a few hundred simulations are necessary for an even cursory
examination of the parameter space and in order for the three 
simulations discussed here to be fiducial they need to have the same
resolution as the rest.}. 
We increased this number by more than an order of
magnitude, adopting for each gas, or
stellar particle a mass of $m_g$=$5 \times 10^4 M_\odot$ and a
softening of 25 pc. The dark matter (DM) particles have a mass of
$m_{\mathrm{DM}}$=$2 \times 10^5 M_\odot$ and a  softening of 50 pc.
We thus have  
in our simulations 2 and 3.5 million particles for the baryons and
DM, respectively. 

The conceptual improvement concerns the initial conditions of the
simulations. In previous works, the progenitors resembled
local 
disk galaxies, except for a higher content of cold gas in the disk.
They consisted of a DM halo, a disk and
sometimes a classical bulge with properties compatible to those of
local galaxies. It is, however, well established that disk
galaxies, except for the cold gas in their disk, have also 
hot gas in their halos (e.g. \citealt{Miller.Bregman.15}).  
To include this, we start off our simulations with 
spherical protogalaxies consisting of DM and hot gas. Before
the merging, a disk forms in each of the progenitors, so that we
witness the merging of two disc proto-galaxies. The two gaseous halos
merge into a single one enveloping the remnant and thus 
halo gas accretes onto the remnant disk all
through the simulation (Sect.~\ref{subsec:general-merger}). Such
gaseous halos exist also in mergings occurring in  
cosmological simulations (e.g. \citealt{Governato.BBMWJSPCWQ.09}),
or in a couple of major merger studies 
\citep{Moster.MSNC.11, Kannan.MFMKS.15}, whose remnants have a $B/T$
ratio between 0.7 and 1., thus linking them to ellipticals with a
small disk. Thus,     
to our knowledge, our study is the first one to include a hot gaseous component
in dynamical simulations of major mergers whose
remnants model realistic spiral galaxies.


\subsection{Code}
\label{subsec:code}

A full discussion of the code used in these
simulations and of their initial conditions is the subject of the next
paper in this series (Rodionov et al., in prep., Paper II). In this and the next
two subsections we only summarise the main information necessary
here. 
 
We use a version of \textsc{gadget3} including gas and its physics
\citep{Springel.Hernquist.02, Springel.05}. DM  
and stars are modeled by N-body particles, and gravity is
calculated with a tree code. The code uses an entropy conserving
density driven formulation of SPH with adaptive smoothing lengths
\citep{Springel.Hernquist.02} and subgrid 
physics \citep{Springel.Hernquist.03}. 

Our feedback, star formation and cooling follow subgrid physics included in
simple recipes given by \cite{Springel.Hernquist.03} and were
  already used in a number of previous works cited in Sect.~\ref{sec:intro}.
  It is beyond the scope of this paper to test other
  subgrid physics. Nevertheless, let us mention 
briefly that \cite{Cox.JPS.06} showed that the mass profiles of the
major merger remnants are robust to substantial changes in subgrid physics
parametrizations. Similarly, \cite{Hopkins.CYH.09} find that the
efficiency with which gas avoids consumption during the merger and can
reform a disk does not depend on the subgrid parametrization, unless
the feedback is very weak.     
Most important, \cite{Hopkins.CHNHM.13} introduced detailed,
explicit models for stellar feedback in 
their very high resolution
major merger simulations and found that in all cases the mass profile
results of explicit 
feedback models are nearly identical to those obtained with the
subgrid physics we use here, except for some second order
differences.  

\subsection{Central, AGN-like feedback}
\label{subsec:AGN}

The code described above has already been used 
a large number of times to test the relevance of major mergers
regarding the formation of disk galaxies, as e.g. in 
\citeauthor{Hopkins.CYH.09} (\citeyear{Hopkins.CYH.09}, and
  references therein), albeit with 
initial conditions that do not include gaseous halos. We also used it 
in a number of our 
simulations, which always include gaseous halos. In those cases, we
obtained merger remnants which are disk 
galaxies with thin and extended disks and realistic spiral arms, but
with one serious drawback which is that the centermost part of the galaxy
had a considerable central concentration, leading to an unrealistically high inner
maximum of the circular velocity curve (Paper II).
Furthermore, this high central mass 
concentration has the disadvantage of prohibiting bar formation, or at
least delaying it beyond the 10 Gyr covered by our simulations. This
is in disagreement with observations, since about 2/3 of local disk galaxies
are barred (see references in Sect.~\ref{sec:intro}).

Such excessive concentrations were also obtained in cosmological simulations
of disk galaxy formation and were
lately addressed by introducing additional feedback in the central
regions, mainly in the form of AGN feedback. Very schematically, in this
picture gas will flow inwards to the central black hole (BH). Feedback is
then calculated as a given fraction of the luminosity which is
radiated by the BH. This energy is distributed to the gas in the central
region in the form of thermal energy, thus preventing excessive
star formation. 

In most studies, the inflow on to the BH is modeled
using the Bondi accretion formalism \citep{Hoyle.Lyttleton.39,
  Bondi.Hoyle.44, Bondi.52}, sometimes limiting it
by the Eddington accretion rate to prevent excessive
accretion. This inflow formalism has two free parameters, the
accretion efficiency and the radiative efficiency
\citep{Springel.DMH.05}. However, the physics of driving
BH-generated outflows is still not well understood
\citep[e.g.][]{Silk.Mamon.12}, and also necessitates resolutions much higher
than what we have here. It has, furthermore, been criticized
by \cite{Hopkins.Quataert.10}, who calculated the accretion directly
from subparsec `resimulations' of the central region of galaxy-scale
simulations.    

Our approach is based on the same physics, but since all we want is to
solve the excessive mass concentration problem and not to model the BH
accurately, we adopted a very straightforward, empirical and
parametric method. As 
gas flows inwards, it will increase the density in the central regions.
As in previous descriptions, we introduce two parameters, a density 
threshold $\rho_{\mathrm{AGN}}$ and a temperature $T_{\mathrm{AGN}}$. More specifically,
at every time step we give internal energy to gas particles whose
local density is larger than the
threshold $\rho_{\mathrm{AGN}}$, by increasing their temperature to
$T_{\mathrm{AGN}}$. This density threshold is chosen so as to ensure that the
chosen particles are located in the centermost region. Furthermore, to
  ensure that the amount of energy that we thus inject is not
  excessive, we test at every step that the 
total energy distributed is below that of the Eddington rate (see e.g.
\citealt{Springel.DMH.05}). We do this in a probabilistic manner,
by setting the probability of a particle receiving internal
energy equal to 1 if the energy to be distributed is less than the
Eddington limit, or, if it is higher than that limit, setting the
probability as the ratio of the Eddington limit to the energy we were
initially to
distribute. A fuller description  
will be given in Paper II, where we describe all our
computational and technical aspects in detail. 
Although very simple, this description includes the essentials sufficient
for our purposes, namely it injects energy in the central regions
  to prevent excessive star formation.
 Compared to cases with no AGN feedback, it leads to 
mass distributions which are less concentrated in the centermost parts,
more realistic circular velocity curves and allows bars to grow.
  
In our scheme there is no single particle representing the 
BH, as any spurious off-centering of such a particle, or its
imperfect correction by 
analytical drag forces, 
could introduce errors  
because the BH mass is so much higher than that of the other particles in
the simulation. 
Nevertheless, it is useful to  
keep track of the evolution of the BH mass ($M_{\mathrm{BH}}$) with time. We do
this by measuring the energy released by our AGN-like feedback as a
function of time and then following the simple formalism described in
\citeauthor{Springel.DMH.05} (\citeyear{Springel.DMH.05}, p. 783). 

The following three questions need to be considered here:

\begin{enumerate}
\item
{\it Does the thus described feedback model lead to unphysical
  results?} Several arguments argue against this:
\begin{itemize}    
\item
The shape of the circular velocity curve, from unrealistic as in
simulations  
without this central feedback, becomes very realistic, well compatible
with observed rotation curves. 
\item
The mass of the BH at the end of the simulation, calculated using
the energy of the feedback, is 
1., 1.4 and 3.3 $\times$ 10$^7 M_\odot$ for our three fiducial
simulations. These are  
reasonable values for disk galaxies. Moreover, we calculated the
velocity dispersion of the central parts, $\sigma$,
directly from our simulation and found that this ($M_{\mathrm{BH}}$, $\sigma$)
pair falls near the
$M_{\mathrm{BH}}$ --- $\sigma$ relation \citep{Ferrarese.Merritt.00,
  Gebhardt.p.00}, within the observational spread. 
\item
Qualitatively, the evolution of $M_{BH}$ during the merging 
is similar to that obtained with other AGN feedback
prescriptions, namely increases much more sharply during the merging
than after it \citep[e.g.][]{Thacker.MWH.14}. 
\item
Last, but not least, we ensure that at all times the amount
of injected energy does not exceed the Eddington limit 
as described above. 
\end{itemize}   

\item
{\it How is this energy distributed?} We followed during the simulations
the locations of the gas particles to which the energy is
deposited. We found that these particles cover a region around the
dynamical center of the halo, i.e. around the position of the halo particle
with the highest local density. The area they cover may vary from one run to
another, but it is always much larger during the merging times. At
such times, their characteristic size is of the order of half to one kpc,
but becomes considerably smaller 
after the merging, when it is of the order of say 
100 pc.  
This increase of
the characteristic size during merging  
reflects the corresponding much larger feedback, 
while the extent during the merging time can be
compared to the circumnuclear region which is very active during a
merging (see e.g. \citealt{Scoville.SSS.91} for the merging system Arp 220). 

\item
{\it Does this feedback affect the simulation results?} Indeed it does, as
expected and required. Namely it changes the shape of the circular velocity curve,
making it compatible with observations and, as a corollary, it allows
the formation of bars, which we know are present in the majority of
nearby galaxies (see references in Sect. 1). 
Moreover, changes in the values of the two adopted parameter 
($\rho_{\mathrm{AGN}}$ and $T_{\mathrm{AGN}}$) should, and do, affect the 
simulation results. This is expected since they change the
shape of the circular velocity curve and thus the bar properties,
which, in turn, influences the dynamics and structure at least in the
inner parts of the disk. These changes are most important in the
central region hosting the bar, but much smaller at larger radii (Paper II).
\end{enumerate}

To summarise, our simple parametric description of the central
feedback has the desired effect without being unphysical. We thus
adopt it for our simulations.   

\subsection{Further information on the code}
\label{subsec:GIZMO}

Making a full comparison of the results of our simulations with those
obtained with other codes, but with the same initial
conditions and the same gas physics and subgrid physics, is beyond
the scope of this paper. We, nevertheless, reran one of our
fiducial simulation (referred to in the following as mdf732, see
Sect.~\ref{subsec:3-examples}) using the \textsc{gizmo} code 
\citep{Hopkins.13, Hopkins.14, Hopkins.15},
after  implementing to it the subgrid physics we use
here. This code is a derivative code of GADGET, using the same MPI
parallelization, domain decomposition, gravity solvers, etc. 
However, in contrast to that of \textsc{gadget}, the SPH version of
\textsc{gizmo} is density independent, and it also includes a more
sophisticated treatment of the artificial viscosity term
\citep{Cullen.Dehnen.10}, as described in \citeauthor{Hopkins.13} 
 (\citeyear{Hopkins.13}, Sect. 3.1). 
It can thus handle better the phase boundaries,
e.g. between the hot gas in the halo and the much cooler gas in the
disk, or the description of cold dense clumps of gas in the
otherwise hot halo. 

The main difference we found between the results of these two codes is in
the hot gaseous halo. This has small gaseous clumps in the
simulation run with \textsc{gadget3}, which are either absent,
or much less prominent in the \textsc{gizmo} run (see also
  \citealt{Torrey.VSSH.12} and \citealt{Hayward.TSHV.14}). 
On the other hand, we find very good agreement when we compare 
the properties of the remnants, e.g. their gaseous or stellar
projected density radial profile, 
their cumulative stellar mass, or the mean tangential velocity of the
stars or gas and the stellar radial and vertical velocity dispersion
profiles.  

The good agreement between the 
properties of these two simulations argues that
the clumps present in the halo of the \textsc{gadget} simulation do
not influence much the behaviour of the 
stellar component. Even the stellar vertical velocity dispersion or
the thickness of the stellar disk are in good agreement, which shows
that these clumps are not sufficiently massive and/or
numerous to heat up noticeably the stellar disk.


\subsection{Initial density profiles of the protogalaxies}
\label{subsec:isolated-dens-prof}

We used idealised initial conditions with two identical 
protogalaxies composed of spherical DM and  
gaseous halos, of masses $M_{DM}$ and $M_{gas}$, respectively, with
$M_{total}/M_{gas}=8$. The initial density of the DM and the gaseous halo is   
 
\begin{equation}
\rho(r)={C~~sech(r/|r_t|)}~~{x^{-\gamma_i}~(x^\eta + 1)^{-(\gamma_{o}-\gamma_{i})/\eta}},
\label{eq:halodens}
\end{equation}  

\noindent
where $r$ is the spherical radius, $x=r/a$ and $a$ and $r_{t}$ are
the scalelength and tapering radius of the halo,
respectively. The constants $\gamma_{i}$,
$\gamma_{o}$ and $\eta$ characterize the shape of the radial density
profile.  
The DM halo has
$\gamma_i=1$, $\gamma_o=3$ and $\eta=1$, while the
gaseous one has $\gamma_i=0$, $\gamma_o=2$ and $\eta=2$ (Beta
model for $\beta=2/3$, e.g. \citealt{Miller.Bregman.15} and references
therein). For both we have $r_{t}$=100 kpc and $a$=14 kpc. The halo
component was built following \cite{McMillan.Dehnen.07}, i.e. with a
\cite{Cuddeford.91} distribution function. 
We also add a small amount of spin to the halo. Let
$f$ be the fraction of particles with a positive sense of rotation. 
In cases with no net rotation $f$=0.5,
while if all particles rotate in the positive (negative) sense 
$f$=1 (-1). The internal energy of the gas was set by requiring hydrostatic
equilibrium (Paper II). 

Stars do not exist in the initial conditions,
but form during the evolution, so that, at the time of the merging,
a stellar and a gaseous disk are present in the progenitor galaxies. 
Thus their density and velocity profiles are set by our initial
  setup for the hot halo gas distribution and its subsequent
  evolution, and not directly by the modeller seeking to match
  e.g. what is measured in nearby galaxies.

\subsection{Three fiducial examples}
\label{subsec:3-examples}

In this paper we discuss three simulations (mdf732, mdf778 and mdf780). 
All three are 1:1 mergers of two identical protogalaxies. We start by
adopting the simple case of coplaner mergers, while other orientations
will be considered in forthcoming studies. 
We just mention briefly here that preliminary results show, as
expected, that the orientations
of the spin axes with respect to the orbital plane influence considerably
the properties of the remnant, and that this may account for the 
spread observed in disk sizes, $B/T$ ratios and gas fractions of nearby galaxies. 

\begin{figure*}
  \centering
  \includegraphics[scale=0.31]{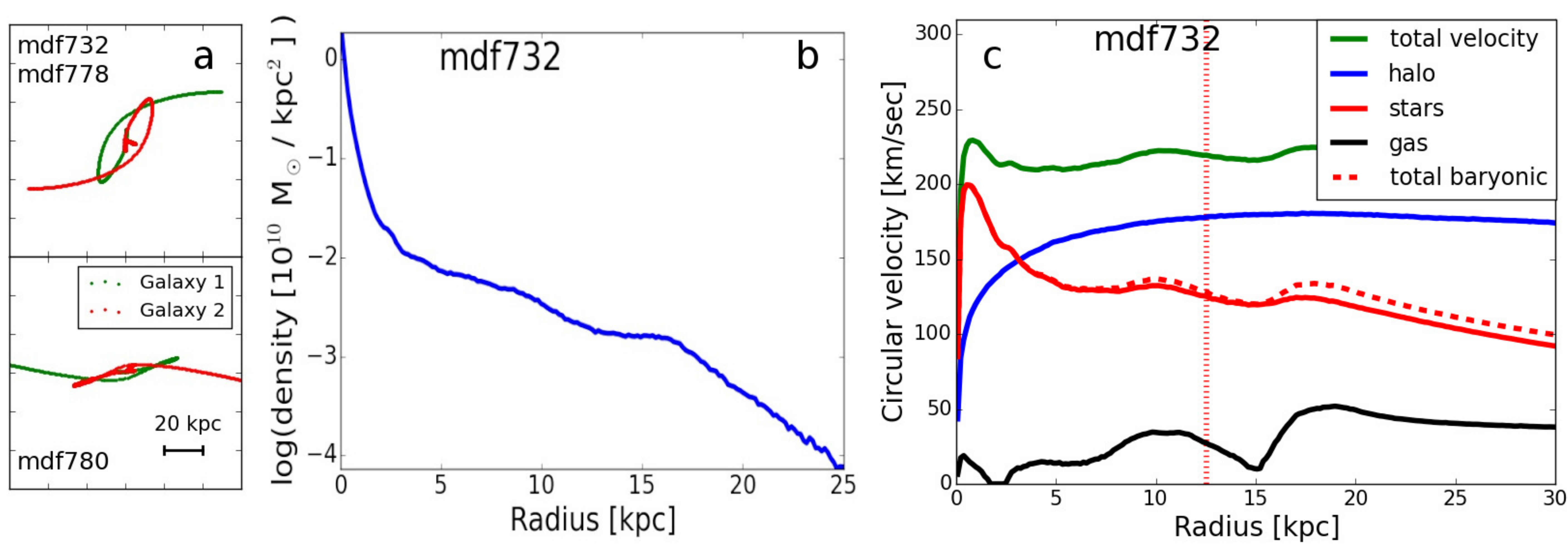}
\caption{a: Orbits of the progenitors before the merging.
    b: Stellar radial surface density profile at $t$=10 Gyr. 
    c: Circular velocity curves at $t$=10 Gyr,
    for the total mass, as well as separately for stars, gas,  
    total baryonic mass and halo. The vertical red dotted line is
    located at 2.2 inner disk scalelengths. 
    }
  \label{fig:vcirc}
\end{figure*}

The orbits of the centers of density of the two protogalaxies,
i.e. the orbits of the location where  their DM density is highest,  
are given in Fig.~\ref{fig:vcirc}a. 
Simulations mdf732 and mdf780 have $f$=0.6 and mdf778 has 0.55. The
adopted AGN parameters are $\rho_{AGN}$=1 $M_\odot/pc^3$ 
for all three cases, and  
$T_{AGN}$=10$^7$ K for mdf732 and mdf778, and  2.5 $\times$ 10$^7$ K for
mdf780. We continued all simulations up to 10 Gyr.

\begin{figure*}
  \centering
  \includegraphics[scale=0.65,angle=-90]{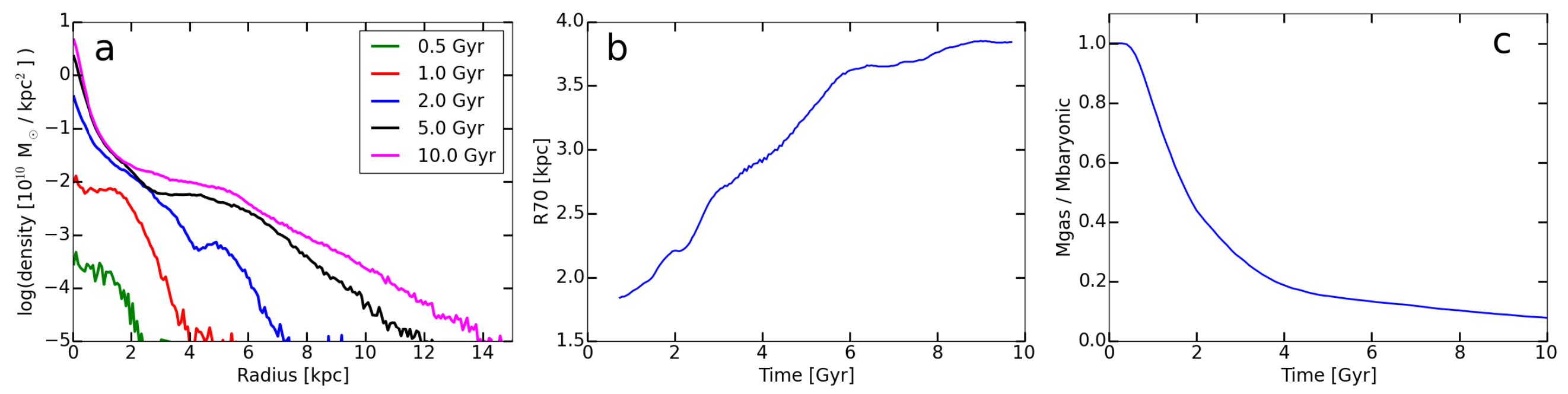}
  \caption{Properties of a simulated galaxy growing in
    isolation. Left: Radial surface density profile for times 0.5 (green), 1
    (red), 2 (blue), 5 (black) and 10 Gyr (violet). Middle:
    Cylindrical radius containing 70\% of the stellar mass. Right: Fraction
    of the gas in the disk ($|{\Delta}z|$$<$0.5) component as a function of time . 
    }
  \label{fig:early-profiles}
\end{figure*}


\section{Results and discussion}
\label{sec:results}

\begin{figure*}
  \centering
  \includegraphics[scale=0.4]{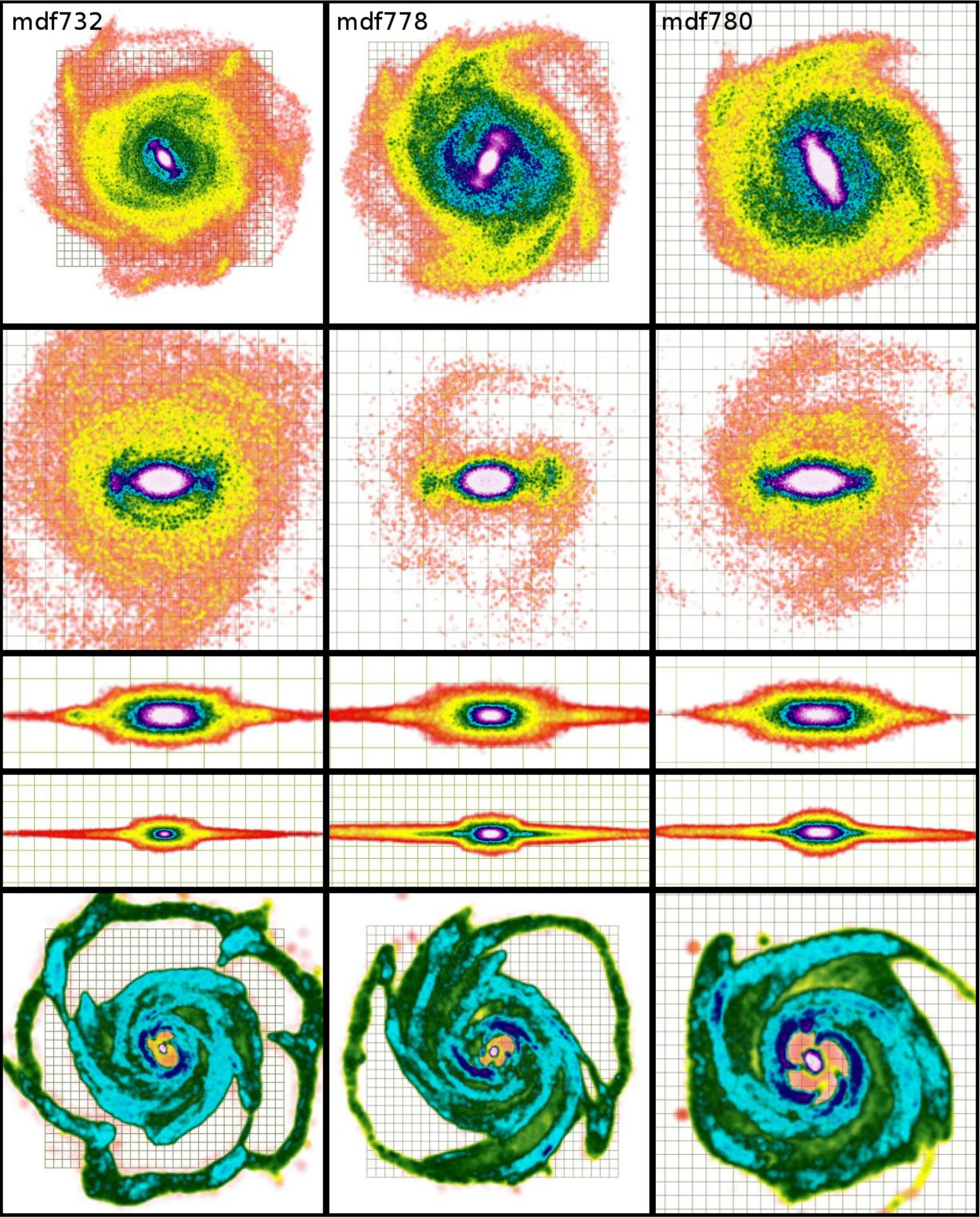}
  \caption{Morphology of the stellar (upper four rows) and gaseous
    (fifth row) disk components of our
    three fiducial simulations at $t$=10 Gyr. The uppermost panels give the 
    face-on view and the second row zooms in 
    the inner regions to highlight the face-on bar morphology. The
    fourth row shows the side-on view of the disk and the third
    row the side-on view of the bar region. The fifth row gives the
    face-on view of the gas distribution. To
    bring out best the features of interest (see text), we choose the
    color coding and linear resolution of each panel separately and 
    include in the background a Cartesian
    grid with cells of 1$\times$1 kpc size, to allow size estimates. In the
    second, third and fourth row the snapshot has been rotated so that
    the bar is along the $x$ axis.}  
  \label{fig:imagest10}
\end{figure*}

\begin{figure*}
  \centering
  \includegraphics[scale=0.23]{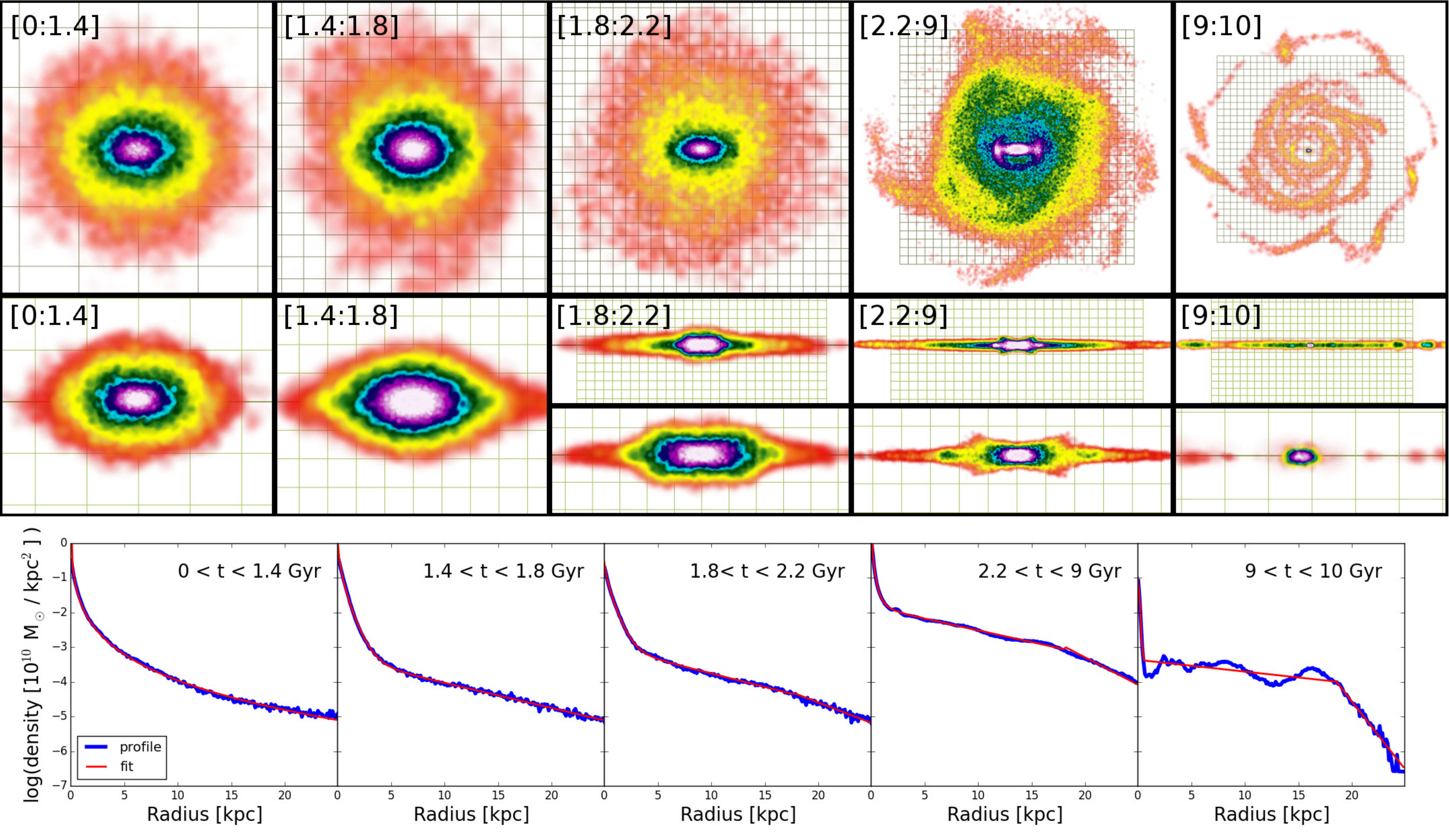}
  \caption{Five different stellar
    populations for mdf732 at $t$=10 Gyr, separated according to 
    their time of birth (see text). Upper panels: Face-on views of
    the five populations. 
    Middle panels: Edge-on views of
    the same populations. For the three youngest populations we
    include also a zoom of the inner parts to show best the
    morphology of the bar region viewed edge-on. For
    color coding and the background Cartesian grid see caption of
    Fig.~\ref{fig:imagest10}.  
    Lower panels: Surface density radial profiles of the five stellar
    populations as obtained from their face-on views (blue lines) and
    the corresponding decompositions (red lines). In the leftmost
    panel we used only one component, of S\'ersic profile. The
    respective birth time ranges are given in 
    the upper right corner of the lower panels.} 
  \label{fig:imagesfivetimes}
\end{figure*}


\subsection{Early evolution of individual isolated galaxies}
\label{subsec:early-isolated}

In previous works (see papers cited in Sect.~\ref{sec:intro}), the
initial conditions consisted of two fully 
developed disk galaxies set on a given trajectory. Each consisted of a
spherical halo, an exponential stellar disk and, in some cases, a
classical bulge and/or a gaseous disk. Their
properties were chosen so as to represent 
nearby galaxies; except for the fraction of gas in the disk,
which was a free parameter varied at will. Thus gas fractions in the
complete range between 0 and 1 were tried, even if this was not necessarily
compatible with e.g. the adopted disk scale length (which was most
times appropriate for low-redshift Milky-Way-like disk galaxies, i.e. 3 to 5
kpc), or with their $B/T$ mass fraction. Although not
ideal, these initial conditions allowed 
previous works to reach interesting conclusions, for example on the
crucial role of the initial gas fraction on the nature of the merger
remnant. This is presumably due to the fact that violent relaxation
occurring during the merging wipes out all small details of the
initial stellar structures.

We chose to start out with very different initial conditions for the
progenitors (Sect.~\ref{sec:sims}). However, before using these initial
conditions in our major merger simulations we made sure
that, when evolved in isolation, their $t$=10 Gyr snapshots gave
reasonable approximations of local disk galaxies, and that during their
early stages of evolution (up to 
say 2 Gyrs) they were compatible with the main properties of disk
galaxies at such times. The latter entails that the disk should be
considerably smaller and more perturbed than the disks in local
isolated galaxies and that the fraction of gas in their disk should be higher
\citep[][etc]{Bouwens.IBBF.04, Ferguson.P.04, Elmegreen.EVFF.05,
  Dahlen.MDFGKR.07, Erb.SSPRA.06, Leroy.WBBBMT.08, Daddi.BWDCDEMROSKS.10,
  Tacconi.P.10, Conselice.MBG.12, Rodrigues.PHRF.12,
  Genzel.p.15}. More information 
on all this will be given elsewhere (Peschken et al., in prep.), but
we summarise below some relevant information.

In our simulations, the protogalaxies at $t$=0 consist only of
spherical distributions of DM and 
gas (Sect.~\ref{subsec:isolated-dens-prof}). From the onset of the
simulation, however, the gas in the halo cools radiatively  
and, getting out of equilibrium, falls inwards. Its density increases
locally and the first stars form. Thus, the progenitors gradually
acquire a disk. This, as expected, is much more perturbed
and lumpy than the more settled exponential disks to which it evolves
at later times. Fig.~\ref{fig:early-profiles}a shows the radial
surface density 
profile at five times during the evolution (0.5, 1, 2, 5 and 10
Gyr). It is clear that the disk grows inside-out, relatively fast
at early times and less so later on. The profiles are initially
exponential-like, but it is not trivial to measure adequately their
scale-length, so, instead, we measure the cylindrical radius $R_{70}$,
containing 70\% of the stellar mass. It is clear that this quantity increases
considerably with time, as expected, particularly at early evolution
stages (Fig.~\ref{fig:early-profiles}b).

Fig.~\ref{fig:early-profiles}c shows the fraction of gas in the disk
component 
measured within $|{\Delta}z|$$<$0.5 kpc. We see that at the relevant times,
i.e. the times corresponding to before the
merging, the gas fractions are in the ball park of 60 to 30\%,
i.e. considerably larger than the corresponding gas fractions in
nearby disk galaxies, in good agreement with observations of both
local and intermediate redshift galaxies.

We do not claim that this is a perfect model of galaxies 
at intermediate redshifts, but it is still a considerable improvement over
those used in previous studies. This improvement is a corollary of the
existence of a hot gaseous halo in our initial conditions, as this
entails a slow formation and evolution of the disk
which becomes gradually more massive and more extended, while
the gas fraction in the disk decreases monotonically with
time.

\subsection{General evolution of a major merger}
\label{subsec:general-merger}

The stellar disks of the progenitor galaxies are destroyed by the
merging and their 
stars concentrate in the central regions of the remnant. As we will
see in the following sections most of them form a classical bulge,
while the remaining, albeit much smaller number of stars, contribute to a
thick disk and its bar. The gaseous disks of the progenitors are also
destroyed, and the gas which they included 
also concentrates in the innermost regions of the
remnant, so that the stars that form from it contribute to the 
classical or disky pseudobulge. 

After the end of the merging, gas continues to
fall in mainly from the halo, but also, although in much smaller
quantities, from the gaseous tails formed during the interaction. Thus a
new disk is gradually formed, which very early on shows 
well-defined spiral arms. This gas accretion continues all through the
simulation.

\subsection{Morphology of merger remnants}
\label{subsec:morphology_res}

Fig.~\ref{fig:imagest10} gives the  morphologies
at the end of the simulations\footnote{Figs. \ref{fig:imagest10}, 
  \ref{fig:imagesfivetimes} and \ref{fig:diskpop1}
were made using the glnemo2 software 
    (http://projets.lam.fr/projects/glnemo2), which, in order to
display morphologies best, uses color coding corresponding to the
maximum of the spatial density along the line of sight.}.
mdf778 has the largest disk extent and mdf780 the shortest.  
Seen face-on, all three galaxies have a clear and
extended spiral structure. Let us denote by $m$ the spiral arm
multiplicity, i.e. the number of arms at any given radius.   
mdf778 is mainly bisymmetric ($m$=2), while mdf732 and
mdf780 are $m$=2 in the inner parts and $m$$>$2 at
larger radii, reaching $m$=5 at the edges of the disk. This increase
of $m$ with distance from the center is in good agreement with  
observations and was explained by \citeauthor{Athanassoula.BP.87}
(\citeyear{Athanassoula.BP.87}, see also \citealt{Athanassoula.88} for
models including accretion). 
These authors applied the swing amplification analytical formalism
(\citealt{Toomre.81}; see also reviews by
\citealt{Athanassoula.84} and \citealt{Dobbs.Baba.14}) to their
  rotation curve decompositions results and pointed out that,
if the spiral structure was due to swing amplification,
the arm multiplicity should increase with distance from the center.  
Thus the spiral structure in mdf732 and mdf780 is
consistent with a swing amplification origin. Providing, however,
tangible proof that these are swing amplification driven spirals is well
beyond the scope of this paper.

mdf732 has an inner ring with a very asymmetric density distribution
along it (Fig.~\ref{fig:imagest10}, second row, where the lower half of the ring is
clearly visible). Its semi-major axis has the same size as that of the bar and
it is elongated in the same direction, both in good agreement with
inner ring observations \citep[e.g.][and references
  therein]{Buta.95}. 
Good examples of such inner rings can be seen in NGC 1433 and NGC 3660. 
mdf778 and mdf780 have m=2 spirals
emanating from the ends of the bar roughly perpendicular to it and
then trailing behind it, again in 
good agreement with what happens in observed local galaxies. 

The bar size in all three simulations (2.5 to 4 kpc semi-major axis)
is only slightly below the average for the 
stellar mass in question, and well within the range of observed values
for external galaxies \citep[lowest panel of Fig. 20
  in][]{Diaz.GSLHE.15}. 
It should be noted that our three models presented here have a similar
stellar mass as the Milky Way, and their barlength closely matches 
that of the Milky Way bar.
To measure the
bar strength, we Fourier decomposed the projected surface density of
the face-on view as in \citeauthor{Diaz.GSLHE.15} and used the maximum
value of the relative $m$=2 Fourier component as the bar strength (see
e.g. \citeauthor{Diaz.GSLHE.15} for more information on this
method). For our three simulations we find values between 0.32 and
0.35, which are in good agreement with
observations. However this comparison is less constraining than the
one concerning the bar length, because the
observed values show a very large spread \citep[second panel of Fig. 20
  in][]{Diaz.GSLHE.15}.  

We can also make meaningful comparisons concerning the bar morphology.
Combining information from three different approaches, namely
simulations following bar formation in isolated disk galaxies, orbital
structure theory, and observations \citeauthor{Athanassoula.05}  
(\citeyear{Athanassoula.05}, see
also \citealt{Athanassoula.15} for a review, and references
therein) came to the conclusion that bars have
a rather complex shape. Namely their outer part is thin both when
seen edge-on and face-on, while their inner part is thick again both
viewed face-on and edge-on. Viewing the bars
in our simulations from different angles, we find that 
they also have the above described morphology, arguing for a proper
dynamical description of bars in our simulations and, most important,
arguing that
the observed bar shapes are compatible with a scenario of disk
formation via major mergers.
 
The face-on bar morphology also agrees well with observations. In
particular, the bar has ansae \citep{Sandage.61} and a barlens component
\citep{Laurikainen.SABHE.14, Athanassoula.LSB.15}. Seen side-on, the latter
is usually referred to as a boxy/peanut/X bulge, but is in fact a part of
the bar, i.e. consists of disk material and forms via disk
instabilities \citep[][for a review]{Athanassoula.15}. 

The morphology of mdf778 shows a number of embedded structures, namely a
bar of radial extent $\sim$3 kpc, an oval of radial extent $\sim$14.5
kpc outside the bar and two $m$=2 sets 
of spirals. The inner set is confined within the oval, while the outer
one emanates from the ends of the oval, extends to larger radii
(maximum $\sim$16.5 kpc) and falls back to the oval thus forming an
outer pseudoring. Similar embedded structures can be seen e.g. in
NGC 1566.   
The outer pseudoring and the oval have the same low pattern speed
($\sim$18.5 km/sec/kpc),
while the inner bar has a much higher pattern speed ($\sim$43
km/sec/kpc). This morphology is very similar to
that of manifold-driven outer spirals and pseudorings
\citep{Romero.GMAGG.06, Athanassoula.RGM.09}. Furthermore,
for manifold theory the pattern speed of the oval and the outer
spirals should be the same, as is the case in mdf778. The ultimate
test, however, can be carried out with the help of the orbits of the
particles in the spirals \citep{Athanassoula.12}. Indeed in any
density wave-based theory the orbits should traverse the arms, staying
longer in the arm than in the interarm. On the contrary, in manifold
theory the orbits of the particles start from one of the Lagrangian
points nearest to the end of the bar (L$_1$ or L$_2$) and then 
follow the arm shape, outlining it, until they reach the opposite side
of the bar. We followed the orbits of particles in the spirals and
found that they stay within the spirals and outline them, 
i.e. these structures are manifold spirals as in the simulations of
\cite{Athanassoula.12}. There is thus 
conclusive evidence for manifold spirals in mdf778.

The gas has, in all three cases, a qualitatively similar morphology to
the stars. Note, however, that, contrary to the stars, the gas has a
minimum in the central region of the galaxy of extent comparable to, but
somewhat smaller than, the bar size. Further in, at a scale of 1 kpc or
less, there is a strong concentration of gas. This morphology is in
good agreement both with observations and with results of previous
more idealised simulations (see e.g. Fig. 2 in
\citealt{Athanassoula.92}). 
 
\subsection{Density and circular velocity profiles of the merger remnants}
\label{subsec:profiles_res}

The total stellar mass at the end of the simulation is around 5
  $\times$ 10$^{10}$ $M_\odot$ in all three simulations, i.e. these
  galaxies  are Milky Way-sized.
Fig.~\ref{fig:vcirc}b shows the azimuthally averaged radial projected
surface density profile of mdf732 at $t$=10 Gyr. Those of mdf778 and
mdf732 are qualitatively the same. The outer parts in all three cases
show a downbending truncation (i.e. of Freeman type II,  
e.g. \citealt{Freeman.70, Erwin.BP.05, 
  Munoz.plus.13}), which is the most common amongst the
truncation types: $\sim$60\% according to \cite{Pohlen.Trujillo.06}, or
42\% according to \cite{Laine.plus.14}. 

Fig.~\ref{fig:vcirc}c shows the circular velocity curves for mdf732 at $t$=10
Gyr, both for the total mass in the galaxy, and for its basic components.
It is calculated directly from the particle masses and positions in the
simulation, without assuming spherical symmetry, contrary to a number
of previous works. It is qualitatively the same for the other two
cases. 
The total curve is fairly flat, with two shallow bumps
due to the spirals. There is also a somewhat higher bump (15 --
20 km/sec) in the central region, due to  the bulge.
The ratio of the velocity due to the baryonic components to the total at 2.2 inner
disk scalelengths is about 0.57 (mdf732), 
0.62 (mdf778) and 0.69 (mdf780) of the total, 
Thus, relying on the \cite{Sackett.97} criterion, mdf732 and mdf778
are clearly submaximal, while mdf780 is also 
submaximal but very near the
borderline with maximum disks. Compared to the
galaxies of the \textsc{DiskMass} survey \citep[][their
  Fig. 2]{Bershady.MVWAS.11}, 
our three models have, for their circular velocity, amongst the
highest fraction of disk mass of the survey, or even
somewhat higher.

\subsection{Circularity parameter}
\label{subsec:circularity}

In a totally cold, perfectly axisymmetric disk, all stars will rotate
with a velocity equal to the circular velocity. Stars in galactic disks,
however, 
have some radial motion, albeit considerably less than their
tangential one. At the other extreme, stars in classical bulges have strongly
non-circular motions with considerable radial velocity components and
with either direct or retrograde rotation. In between the two, there
are stars in structures such as bars, ovals or lenses, whose motion is
neither as circular as that of disk stars, nor as far from circular as
that of classical bulge stars can be. 

These different kinematics are reflected in the different values of the
normalised angular momentum of a stellar orbit, known as the
circularity parameter $\varepsilon$=$J_z/J_{circ}$, where $J_z$ is the 
$z$ component of the angular momentum and $J_{circ}$ is the angular
momentum of the circular orbit of the same energy
\citep[e.g.][]{Abadi.NSE.03, Aumer.White.13}.    
Fig.~\ref{fig:Jevol}a shows an example (red line) of a histogram of
the number of particles as a function of their circularity 
for a snapshot with only a classical bulge and a disk component. 
This simulation is in all aspects identical to mdf732, except that no
AGN feedback has been included (see Sect.~\ref{subsec:AGN} and
Paper II). The classical bulge and the disk are clearly 
distinct and it is possible to distinguish disk from bulge stars
simply by introducing a separation limit to their circularity value
($\varepsilon_{lim}$), which, 
as can be seen from Fig.~\ref{fig:Jevol}a is situated around 0.5.     
We tested this kinematic way of distinguishing disk from classical
bulge stars by viewing the spatial distribution of the two thus
obtained populations separately and from various viewing angles 
and made sure that the disk is indeed adequately distinguished 
from the classical bulge component. After some trials, we found that 
$\varepsilon_{lim}$=0.5 gave quite satisfactory results and we adopted it. 
The stars that this method assigns to the disk(bulge)
will hereafter be called `kinematic disk(bulge) stars'. 

A related, very useful quantity is the circularity at birth
($\varepsilon_{birth}$). As expected, this is very near 1 for stars
born in the disk, but not for stars born in the
bulge. Fig.~\ref{fig:Jevol}b shows the 
fraction of stars of a given birth time which are born with
$\varepsilon_{birth}>\varepsilon_{lim}$, 
as a function of their birth time, for mdf732 and as calculated for
$\varepsilon_{lim}$=0.5 (red curve). Calculations using
$\varepsilon_{lim}$=0.6 and 0.7 give 
quasi-identical results. Note that during the 
merging $J_{circ}$ can not be adequately defined, so the curve 
includes only times after 1.55 Gyr. 
This curve shows considerable 
structure which will be discussed in the next subsection.

\subsection{Morphology and kinematics for populations of different ages}
\label{subsec:agesplit_res}

\begin{figure*}
  \centering
   \includegraphics[scale=0.21]{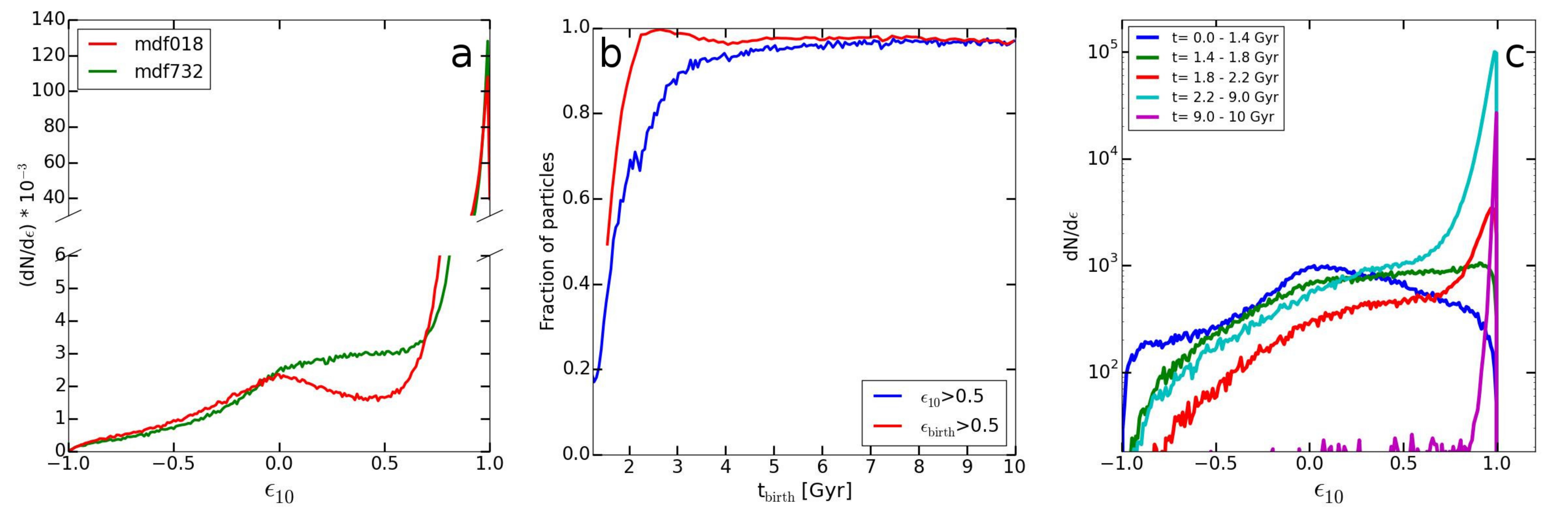}
\caption{Identifying disk stars kinematically. Left: Distribution of the
  number of particles as a function of their circularity
  parameter for two snapshots. In one of these (red) the stars are basically
  either in a classical bulge, or in a disk. In the other (green) it is clear
  that there is a further component. Middle: fraction of stars with 
  $\varepsilon$$>$0.5 at birth (red) and at 10 Gyr (blue).
    Right: Distribution at $t=10$ Gyr of particles as a function of their
    circularity parameter $\varepsilon_{10}$,  
given separately for the stars in each of the five time of birth
brackets as in Figs.~\ref{fig:imagesfivetimes}. 
The adopted bin size in both the rightmost and the leftmost panel is
${\Delta}\varepsilon$=0.01.   
    }
\label{fig:Jevol}
\end{figure*}

It is expected that landmark times, such as the merging time, should set
clear differences between the stellar populations born before and
after them and we want to check whether this is indeed the case in our
simulations and what the corresponding differences are. We will assess
this by examining the $t$=10 Gyr snapshot of mdf732, to which all this
subsection will pertain.

An obvious choice for the first landmark time is the merging
time. In major mergers, however,  
there is no clearly and uniquely defined merging time and it is better
to define a merging period, or time range. Furthermore, we
want to avoid eye estimates of such times, because they
can be biased and are not reproducible. We thus define the
{\bf b}eginning of the {\bf m}erging period as the earliest time
after which the 
distance between the centers of the two merging protogalaxies is
always smaller than 1 kpc ($t_{bm}$). Although arbitrary, this time has the 
advantage of being clearly defined and reproducible. We define as the
center of each galaxy what is often referred to as the center of
density of its halo, calculated from the positions of the
halo particles with the highest local density (as defined by the
distance to their nearest neighbours, e.g. \citealt{Casertano.Hut.85}). 
We measure the distance between the two progenitor centers
and find the earliest time after which this distance stays smaller than
1 kpc. For mdf732 it is around 1.4 Gyr. 

As a second landmark time, hereafter $t_{bd}$, we chose a time
associated with the {\bf b}eginning of the {\bf d}isc formation and,
more specifically, the time beyond 
which the vast majority of stars are born with disk kinematics. We
estimate this   
by using $\varepsilon_{birth}$.  
Fig.~\ref{fig:Jevol}b
(red line, calculated for $\varepsilon_{lim}$=0.5) shows that
the fraction of kinematic disk stars
increases strongly with time of birth for up to 2.2 Gyr  
and that about 95\% of stars born after this time have disk kinematics. 
We thus adopt $t_{bd}$=2.2 Gyr, after verifying that this value holds also
for other reasonable values of $\varepsilon_{lim}$.
We also introduced a third time  at $t_{by}$=9 Gyr, which is somewhat
arbitrary and not a landmark time, but is nevertheless
useful because it sets a time such that any stars born after that can
be considered {\bf y}oung and thus 
allows us to focus on the distribution and kinematics of the youngest
stars.    
For our snapshot, which is at 10 Gyr, the ages of the stars born at
$t_{bm}$, $t_{bd}$ and $t_{by}$ 
are 8.6, 7.8 and 1 Gyr, respectively.

We can thus define five time intervals: namely [0,$t_{bm}$],
  [$t_{bm}$,$t_{hm}$], 
[$t_{hm}$,$t_{bd}$], [$t_{bd}$,$t_{by}$] and [$t_{by}$,10] where  
$t_{hm}=0.5(t_{bm}+t_{bd})$.
We then separate the stars in 
five groups according to which of the above time ranges they were born
in, i.e. according to their age. We thus get five separate
populations, and their respective number of stellar particles is 99369, 
109035, 83960,  713081  and 47376. Their face-on and
edge-on views are shown in Fig.~\ref{fig:imagesfivetimes}. 
Corresponding
kinematic information, separately for each of these five age groups, is given in 
Fig.~\ref{fig:Jevol}c, which shows the distribution of stars as a
function of their circularity at 10 Gyr ($\varepsilon_{10}$). 
Examining this
morphological and kinematic information together, 
we find a number of important results:

\begin{itemize}
\item
Stars born before the beginning of the merging period, i.e. that are
older than 8.6 Gyr, (leftmost panels of
Fig.~\ref{fig:imagesfivetimes}) are concentrated in the innermost  
couple of kpc and their spatial distribution and radial density
profile are that of a triaxial classical bulge. They
are the oldest stars in the galaxy and experienced violent relaxation
due to the strong evolution of the potential during the merging,
thus explaining their very steep 
radial projected density profiles \citep{Lynden-Bell.67}. Thus 
major mergers provide a mechanism for
classical bulge formation. Some of these stars
rotate prograde and others retrograde with respect to the disk, 
as expected for a classical bulge
(Fig.~\ref{fig:Jevol}c).  
There are, however, considerably more prograde than retrograde stars,
i.e. the bulge has internal rotation. 
This is, at least partly, due to the bar, 
which can make the bulge 
more oblate \citep{Athanassoula.Misiriotis.02} and give it some
spin \citep{Athanassoula.03, Saha.MVG.12}. 

\item
Stars born during the first half of the merging period (second column
of panels in Fig.~\ref{fig:imagesfivetimes}), i.e. with ages
between 8.2 and 8.6 Gyr, have a density
distribution similar to that of stars born before the merging,
although more extended and more 
flattened. The outermost isodensities of the edge-on view are cusped,
as one would expect from a relatively light underlying thick disk.
This is corroborated from their angular momentum distribution (green
line in Fig.~\ref{fig:Jevol}c), which shows a considerable 
contribution of 
stars with $\varepsilon_{10}$ near but somewhat less than 1. 

\item
Stars born during the second half of the merging period (third column
of panels) have a very
different density distribution from that of the two previous groups. 
They trace in the outer parts a disk extending well beyond the
bulge region and in the inner parts a bar with  
a vertically thick boxy bulge \citep{Athanassoula.05}. Their
circularity distribution (Fig.~\ref{fig:Jevol}c, red line) also shows the
existence of two components, a fast 
rotating component, i.e. a disk; and a slower component rotating
with an average $\varepsilon_{10}$ around 0.25.
Checking out the spatial
distribution of  
the stars in the latter component we found that it is the
bar (see also Sect.~\ref{subsec:disk-vs-nondisk}). Thus the stars born
in this time range contribute partly 
to the near-axisymmetric disk and partly to the bar.

\item

The stars within the two last age ranges (younger than 7.8 Gyr)
were born during the disk formation era  
and are thus part of 
what is commonly referred to as the disk population. This disk is
extended and quite thin. Except for the
axisymmetric disk, the stars in the fourth age bracket contribute
also to the spirals and to the bar. For this group of stars, the ansae at the
extremities of the bar are clearly visible, and at smaller radii,
there is an X-like edge-on shape, as in many barred galaxies. 
Note that both the disk and the bar are vertically thinner than the
corresponding structures in the third age bracket. 

On the other hand, the stars which are younger than 1 Gyr do 
not participate in the bar structure, but are heavily concentrated in
the spiral and ring structures, as well as in an innermost very thin
structure, which can be called a disky pseudobulge
\citep{Kormendy.Kennicutt.04, Athanassoula.05}. The latter
may well exist in the third and fourth age brackets as well, but is
less easily discernible in plots as in Fig.~\ref{fig:imagesfivetimes}, 
because of the strong bar contribution to the central parts. 

The kinematics of these two youngest age ranges
(Fig.~\ref{fig:Jevol}c) corroborate the above. In particular, they
show that the stars of the fourth age range contribute partly to the
disk and partly to the bar (see also
Sect.~\ref{subsec:disk-vs-nondisk}). On the other hand, the stars in
the youngest age bracket contribute essentially only to
the disk, and more specifically (Fig.~\ref{fig:imagesfivetimes}) to
its spirals, rings and the disky pseudobulge. 

\end{itemize} 

In the lower panels of Fig.~\ref{fig:imagesfivetimes} we show the
projected density radial profiles of each of the five stellar
populations, obtained from their face-on views. The oldest population
is well fitted by a single S\'ersic component with an index
between 4.5 and 6, i.e. corresponding to a classical bulge
\citep{Kormendy.Kennicutt.04, Drory.Fisher.07}. At the other extreme, the
youngest population can be fitted by three exponential disks, the
innermost one corresponding to a disky pseudobulge, and the two next
ones to the inner and outer disks, respectively. Thus, in general, the
classical-bulge-to-total stellar mass ratio 
decreases with the age of the population from 100\% to 0\%.

\subsection{Coupling kinematics and morphology to identify disk stars}
\label{subsec:disk-vs-nondisk}

\begin{figure*}
  \centering
   \includegraphics[scale=0.25]{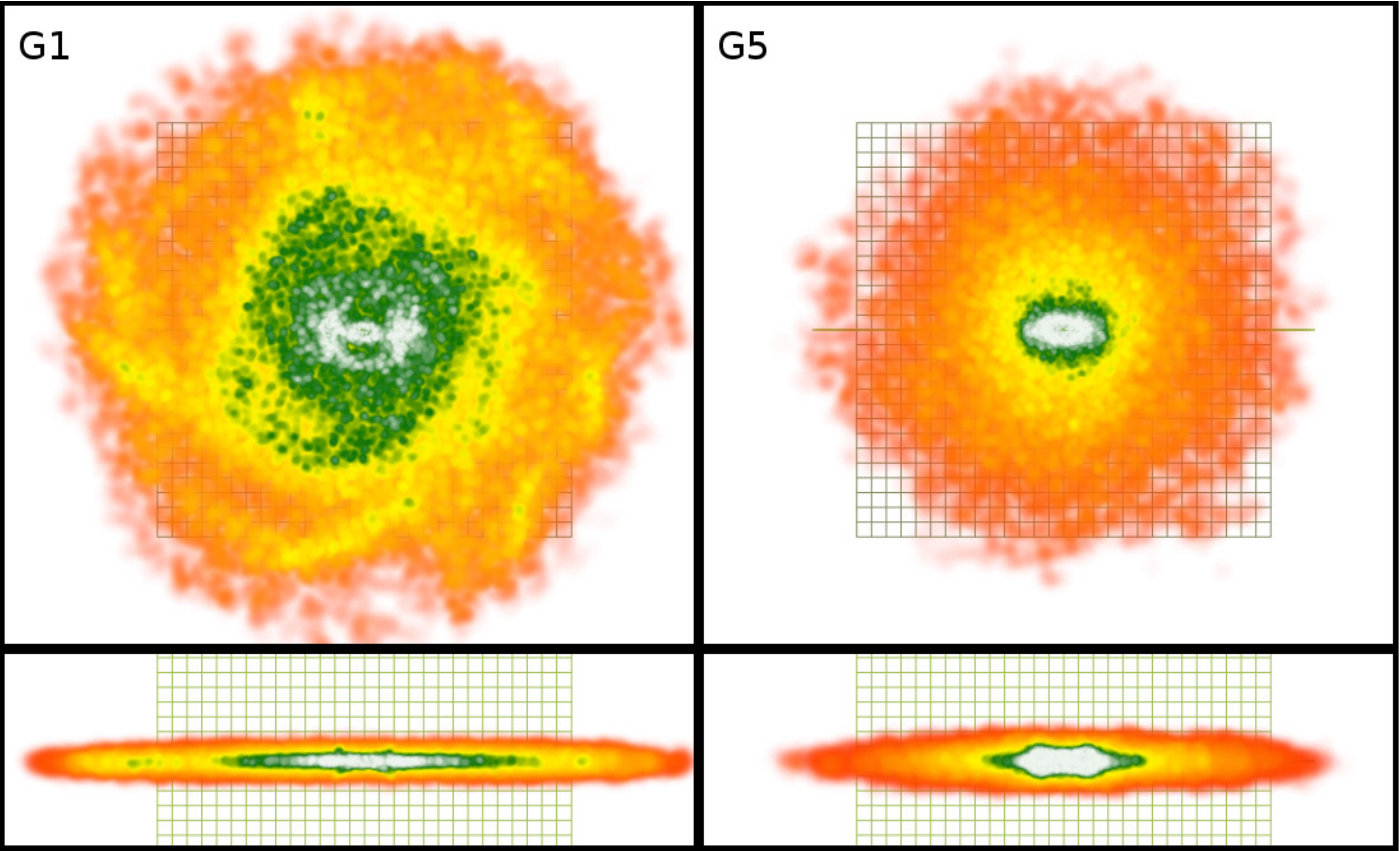}
\caption{Components G1 and G5 at $t$=10 Gyr. Face-on  (upper
  subpanels) and edge-on (lower subpanels) views of group G1 (left)
  and G5 (right) in mdf732 at 10 Gyr. All four
  subpanels have the same linear scale. The color coding was chosen
  so as to bring out best the morphological features of interest
  here. 
  }
\label{fig:diskpop1}
\end{figure*}

\begin{table*}
\caption[]{Basic properties of the eight groups}
\centering
\begin{tabular}{|c|c|c|c|c|}

  \hline
\noalign{\smallskip}
\noalign{\smallskip}
  Group & Birth time & $\varepsilon_{birth}$, or $\varepsilon_{bd}$ & $\varepsilon_{10}$ &
  Fraction \\
\noalign{\smallskip}
\noalign{\smallskip}
  \hline
\noalign{\smallskip}
\noalign{\smallskip}
  G1 & t$_{birth}>t_{bd}$ & $\varepsilon_{birth}>\varepsilon_{lim}$ &
  $\varepsilon_{10}>\varepsilon_{lim}$ & 0.660 \\
  G2 & t$_{birth}>t_{bd}$ & $\varepsilon_{birth}>\varepsilon_{lim}$ &
  $\varepsilon_{10}<\varepsilon_{lim}$ & 0.046 \\ 
  G3 & t$_{birth}>t_{bd}$ & $\varepsilon_{birth}<\varepsilon_{lim}$ &
  $\varepsilon_{10}>\varepsilon_{lim}$ & 0.003 \\
  G4 & t$_{birth}>t_{bd}$ & $\varepsilon_{birth}<\varepsilon_{lim}$ &
  $\varepsilon_{10}<\varepsilon_{lim}$ & 0.013 \\
  G5 & t$_{birth}<t_{bd}$ & $\varepsilon_{bd}>\varepsilon_{lim}$ &
  $\varepsilon_{10}>\varepsilon_{lim}$ & 0.091 \\
  G6 & t$_{birth}<t_{bd}$ & $\varepsilon_{bd}>\varepsilon_{lim}$ &
  $\varepsilon_{10}<\varepsilon_{lim}$ & 0.040 \\
  G7 & t$_{birth}<t_{bd}$ & $\varepsilon_{bd}<\varepsilon_{lim}$ &
  $\varepsilon_{10}>\varepsilon_{lim}$ & 0.024 \\
  G8 & t$_{birth}<t_{bd}$ & $\varepsilon_{bd}<\varepsilon_{lim}$ &
  $\varepsilon_{10}<\varepsilon_{lim}$ & 0.123 \\
  \noalign{\smallskip}
  \hline
\end{tabular}
\label{table:G}
\end{table*}

In Sect.~\ref{subsec:circularity} we discussed a simple way of
distinguishing the classical bulge population from that of the disk for
snapshots with only these two components. We now extend this 
to snapshots with bars and/or ovals and apply it to mdf732 at $t$=10. 
The corresponding circularity histogram is given 
in Fig.~\ref{fig:Jevol}a (green line).
As expected, the distribution is now much more complex. Namely
there is clearly additional material between the disk and the bulge
components and applying a simple separating criterion to $\varepsilon$
would not suffice if an important fraction of stars 
is in the bar component. Independent of their $\varepsilon$, bar stars
should be considered as part of the disk 
component, since the bar forms from a disk instability which
rearranges the disk material. We therefore conclude that not only all
stars with $\varepsilon>\varepsilon_{lim}$ belong to the disk, but also a
considerable fraction of those with $\varepsilon<\varepsilon_{lim}$.   
Thus a more elaborate analysis will be necessary for
galaxies with bars.

We therefore divide stars into groups, depending on three
physical properties, namely  
whether their time of birth is before or after $t_{bd}$, and on
whether their
kinematics at birth and at 10 Gyr are disk-like or bulge-like. 
The latter two are ensured by testing whether their circularity at
birth and at 10 Gyr ($\varepsilon_{10}$) is bigger or smaller than
$\varepsilon_{lim}$. In this section, as in 
Sect.~\ref{subsec:circularity}, we describe results obtained with  
$\varepsilon_{lim}$=0.5. We tried, however, also other values and
reported some of the corresponding results in Sect.~\ref{subsec:B/T}.

This division creates in total eight groups of 
stars, described in Table~\ref{table:G}. Groups G1 to G4 include all
stars born after $t_{bd}$, while 
groups G5 to G8 all stars born before $t_{bd}$. We viewed each group
separately in 3D from different angles 
to assess its shape and morphology and we also made projected surface
density profiles, both radial and vertical and for different
cuts. Whenever a bar is present in the group we
calculated its position angle (PA) and, comparing the PA found for all
groups, we found that they all agree to within 
5$^{\circ}$, i.e. to within the accuracy
of the estimates. We now consider all this information together.

Group G1 
consists of stars with disk kinematics, both at birth and at 10
Gyr. Fig.~\ref{fig:diskpop1} (leftmost panels) reveals a thin
disk with a strong global spiral structure and a rather weak bar. Its 
projected density profile (not shown) is exponential with a small
central bump due to 
the bar. It is thus clearly a disk population and comprises 66\% of
the stars. 

Stars in group G2 
were born with disk
kinematics, and by 10 Gyr evolved to non-disk kinematics. 
Morphology and photometry of G2 show that ~95\% of
G2 stars form a bar and the remaining 5\% a disk component around it.
The bar in G2 has roughly the same orientation and
length as that of G1, but is considerably fatter horizontally. Seen
edge-on, it has a clear boxy/peanut structure. Most of the stars in G2
were born very soon after $t_{bd}$, while those of G1, G3 and G4 are
more evenly spread between $t_{bd}$ and 10 Gyr. Thus the stars in G2
are, on average, older than the stars in other groups. Their
circularity argues that they were born in the axisymmetric disk and
were trapped by the bar as 
this grew. G2 comprises 4.56\% of the stars.  
 
G3 has disk kinematics at 10 Gyr, but not at birth, while G4 does not
have disk kinematics either at 10 Gyr or at birth.
The stars in those two groups are concentrated in an 
inner triaxial object of size roughly 1.4:0.7:0.5,
whose PA is, within the errors,
the same as that of G1 and G2. Thus it can be considered as an inner
part of the bar, presumably part of its barlens component
\citep{Laurikainen.SABHE.14, Athanassoula.LSB.15}. Together, G3 and G4
comprise only 1.65\% of the stars.  
 
Groups G5 to G8 consist of stars with $t_{birth} < t_{bd}$. For most
of these stars it is not possible to calculate $\varepsilon_{birth}$ because
the center of the remnant at such times is not well 
defined. Instead, we calculate $\varepsilon$ at $t_{bd}$
($\varepsilon_{bd}$), because, although this is early on in the
evolution, it is sufficiently late for the   
center of the remnant to be clearly identified. 

The stars of group G5 
constitute at 10 Gyr a thick
disk (Fig.~\ref{fig:diskpop1}) with no spirals and with a bar of
roughly the same length and PA as that of G1, but much
fatter in the plane and with no ansae. 
This thick disk comprises 9.08\% of the stars. 

Stars in group G6 
basically compose a bar,
similar in shape and outline to that of G5, while a few of them are in a
thick disk, similar to that of G5. They comprise 3.99\% of the stars. 

Groups G7 
has too few stars for us to be able to classify. 
It comprises 2.37\% of the stars. 

Most of the stars in group G8 
belong to a flattened bulge, as argued by morphology,
photometry and kinematics together. G8 isophotes
have a bar like deformation in their central part,
analogous to the `halobar' found in the central parts of
halos \citep{Colin.VK.06, Athanassoula.07}. G8 comprises 12.3\% of the
stars.  

The above procedure distinguishes the disk from the
classical bulge component. 
Groups G1, G2, G5, G6 are unambiguously linked to the disk, and this most
probably is true also for G3 and G4. Indeed, they contribute mainly to
the bar and, more specifically, to its inner parts.
Most of group G8 is
linked to the classical bulge. G7 is presumably linked
to the bar, i.e. it is a disk population, but this is not as sure as for the other,
above-mentioned disk   
components. Nevertheless, the uncertainty this entails is very small,
~2 -- 3\%. 

Note also that the stars in the
thick disk are older than those in the thin one, as they were born
earlier in the simulation and in a rather restricted time interval,
roughly between 1.4 and
2.2 Gyr, i.e. at $t$=10 Gyr they have an age between 8.6 and 7.8 Gyr.
Thus the oldest stars in this model are in the
classical bulge, followed by those in the thick disk, while
the youngest are in the thin disk.

\begin{table*}
\centering
   \caption[]{Classical bulge to total stellar mass ratio, calculated
   in four different ways. }
\begin{tabular}{|c|c|ccc|ccc|c|c|}
\hline
\noalign{\smallskip}
\noalign{\smallskip}
  run & $t_{bd}$ & G8 (0.5) & G8 (0.6) & G8 (0.7) & G8 + G7 (0.5) & G8
  + G7 (0.6) & G8 + G7 (0.7) & Hopkins 2009 & Surface density \\
 \noalign{\smallskip}
\noalign{\smallskip}
\hline
 \noalign{\smallskip}
\noalign{\smallskip}
mdf732 & 2.2 & 0.12 & 0.14 & 0.16 & 0.15 & 0.16 & 0.18 & 0.18 & 0.09 --
0.15 \\
mdf778 & 2.2 & 0.11 & 0.13 & 0.15 & 0.13 & 0.15 & 0.17 & 0.19 & 0.10 --
0.18 \\
mdf780 & 2.8 & 0.18 & 0.21 & 0.24 & 0.22 & 0.24 & 0.27 & 0.25 & 0.09 --
0.11 \\
 \noalign{\smallskip}
\hline
\end{tabular}
\label{tab:bulge-to-total}
\end{table*}

\subsection{B/T mass ratio}
\label{subsec:B/T}

In Table~\ref{tab:bulge-to-total} we give various estimates of the $B/T$
ratio in our three fiducial simulations at $t$=10 Gyr, as well as some
close upper 
limits. Columns 1 and 2 give the run number and $t_{bd}$,
respectively. Columns 3 to 8 give results for the method described  
in Sect.~\ref{subsec:disk-vs-nondisk}, where we extended the simple
kinematic decomposition 
of the stars into a disk and a bulge component, to cases with a
bar. There we identified most of group G8 as the classical bulge component
and the corresponding $B/T$ values are given in columns 3 to 5 of
Table~\ref{tab:bulge-to-total}. We were, 
however, unable to safely identify whether G7 should be considered as
a disk or a classical bulge component, so we will for safety also
include an estimate based on the sum of the two components, as an
upper limit (columns 6 to 8). We applied this with
$\varepsilon_{lim}$=0.5 (columns 3 and 6), the value 
we have found to be more appropriate (Sect.~\ref{subsec:circularity})
and \ref{subsec:disk-vs-nondisk}, but also with 
$\varepsilon_{lim}$=0.6 (columns 4 and 7) and even
$\varepsilon_{lim}$=0.7 (columns 5 and 8), the last two,
and particularly the last one, being more like upper limits.   
As expected, the smallest values are when identifying  G8 to the
classical bulge and using
$\varepsilon_{lim}$=0.5, while the largest are the upper limits
obtained when identifying G7
and G8 together to the classical bulge component and using
$\varepsilon_{lim}$=0.7. It is, however, very reassuring that
the differences are small, showing that the upper limits are close to
the most probable values. 

\cite{Hopkins.CYH.09} introduced a different, much simpler and more
straightforward method to obtain an estimate of $B/T$. Namely, they
assume that the bulge has no global rotation and that it is the only
component that includes negative $\varepsilon$ values. Under these
assumptions the contribution of the bulge is equal to twice the mass
of particles that have negative velocities. This was introduced for
disk galaxies with no bars, where the two above assumptions are very
reasonable to make. However, as discussed in Sect.~\ref{subsec:agesplit_res} 
and the references therein, bars, although part of the disk
population, do not have disk kinematics. Furthermore, they can
transmit angular momentum to the classical bulge, so that for strongly
barred
galaxies the estimate from this method can be considered as approximate. 
We apply it to our simulations and find values to within 10\% of our
upper limits (Table~\ref{tab:bulge-to-total}). It is useful to have
established this agreement, because the \citeauthor{Hopkins.CYH.09}
method is easy to apply and was used in a large number of
previous cases.

To get a third, independent estimate of $B/T$, we used a decomposition of
the radial projected density profile, obtained by
  averaging the density in cylindrical annuli in the face-on view of
  the disk. This decomposition is similar to the 1D decompositions
used by observers for the radial luminosity profiles and, although 1D, it
is not straightforward, since the innermost regions may include either
both a classical and a disky pseudobulge, or one of the two
only. Depending on exactly how the 
decompositions are done, we find values for mdf732 between  
0.09 and 0.15, for mdf778 between 0.10 and 0.18 and for mdf780 between 
0.09 and 0.11, i.e. somewhat lower than, but in good agreement with
the kinematic estimates. 

The values of the $B/T$ ratio we find here are considerably smaller than
those found in previous simulations with no hot gas in the halo. 
Such simulations are well
summarized in
\cite{Hopkins.CYH.09}, where, out of several hundred simulations, only five 
have $B/T$$\le$0.2, and, moreover, these five have 1:8-1:10 progenitor
mass ratios. On the contrary, our $B/T$ values are compatible with those
of observed spiral galaxies and do not exclude that such galaxies were formed
from major mergers. This is a big improvement over past works and is due to
the existence of a hot gaseous halo in the progenitors, which is
carried over in the merger remnant. Indeed, as we
showed in the previous subsections, the mass of the
classical bulge (B) is roughly set by the number of stars that formed
before the merging. On the other hand the mass of the disk is mainly due to
stars that formed after the merging. Due to the gaseous halo,
the accretion of gas on the disk continues well after the merging, up to
the end of the simulation and presumably well after it. It thus leads to a
more extensive and more massive disk than what would be found in the
absence of a gaseous halo,
and, therefore, smaller $B/T$ values. Comparisons for more simulations
and the effect of various progenitor properties and orbital parameters
on the $B/T$ values will be given elsewhere.

Observations show that there are spirals, in particular
relatively small late types, with no, or hardly any, classical bulge
\citep{Kormendy.DBC.10}. Our own Galaxy has a 
classical bulge of low mass
\citep[e.g.][]{Shen.RKHDK.10, Ness.p.13a, Ness.p.13b}, but it is
premature to state that it has no classical bulge at all.
Our simulations produced considerably lower $B/T$ values than previous
ones considering major mergers. So far, however, none of our
simulations gave results 
compatible with a total lack of classical bulge. This is a general
problem of all disk galaxy formation studies and may be
solved by changing the feedback recipes, as attempted in cosmological
simulations \citep[e.g.][]{Brook.SGRWQ.12,
  Brooks.Christensen.15}. Here let us simply note that major mergers
should be rarer in low density 
environments, thus if bulgeless disks are found solely, or
predominantly, in such environments their hosts may well not be major
merger remnants. Moreover, our aim here
is to test whether major mergers may produce disk galaxies,
but certainly not to show that they are the 
only way of making them. Thus it is not necessary for our scenario 
to be able to form disk galaxies with no classical bulge at all. .

\subsection{The thick disk component}
\label{subsec:thick}


When examining the group G5 in mdf732 at 10 Gyr
(Sect.~\ref{subsec:disk-vs-nondisk}), we 
found that its stars form a thick disk (Fig.~\ref{fig:diskpop1}). We
will now discuss its properties further and compare them to those of
observed thick disks. Such a comparison can only be approximate
because other groups may also host
some thick disk stars. For example, 
thick disk formation will not stop abruptly at time $t_{bd}$, but will
continue at subsequent times, albeit presumably at a tapered rate.
Moreover, some stars born in the thin disk will be vertically heated,
e.g. by the spirals, and thus contribute to the thick disk. 
Yet it makes sense to use G5 for comparisons with observations, because it is 
defined using physical criteria (Sect.~\ref{subsec:disk-vs-nondisk})
and, furthermore, it is a main,  
presumably {\it the} main, contributor to the thick disk. We will thus
use here group G5 as a proxy, and loosely refer to it as the thick disk
component. Note also that
even in observations there is more than one way to define the thick
disk. Thus \cite{Yoachim.Dalcanton.06},  
\cite{Comeron.thick-global-1.11, Comeron.thick-global-2.12}
and \cite{Streich.P.15} use different methods and/or different fitting
functions, thus introducing considerable differences in the results.

In the disk formation scenario which we consider, the thick disk forms
naturally, in agreement with the observational claim of thick disk 
(near-)ubiquity (\citealt{Yoachim.Dalcanton.06, Comeron.thick-4244.11},
but see also \citealt{Streich.P.15}).

Contrary to the thin disk, the simulated thick disk has no spirals, or
then of such low amplitude 
that they are not discernible by eye. This should be linked to the fact
that both density wave \citep{Lin.67} and swing amplification theories
\citep{Toomre.81} show that hotter and thicker disks will harbour
lower amplitude spirals. A similar result presumably holds also 
for a manifold origin, because it is more difficult to confine a
hot stellar population than a cold one (see also
\citealt{Romero.GMAGG.06}), but no specific quantitative study has yet
been made.

By definition, the stars in
group G1 were born after $t_{bd}$, while those in group G5 before
that. Thus the stars in the thick disk are older than the 
stars in the thin disk, in agreement with observations
\citep[e.g.][]{Mould.05, Yoachim.Dalcanton.08, Comeron.SJLY.15}. 
   
The bar in the thick disk has the same length and orientation
as that of the thin one, but is much thicker in the disk plane than that of
G1, as would be expected from previous work
(\citealt{Athanassoula.83}, see also \citealt{Athanassoula.03}),
where it was shown that the bar is thicker in the disk plane 
in the case of hotter disks. Note also that the bar in the thick disk has no
ansae and, viewed side-on, it has a boxy/peanut bulge with a considerably
larger vertical extent than that of the thin bar. According to this
scenario, and since the thin and the thick disks 
co-exist, the bar component in observed galaxies will include
contributions from both thin and thick disks, so that various parts of the bar may
have different mean stellar ages.

Observations \citep{Yoachim.Dalcanton.06, Comeron.thick-global-2.12}
show that the ratio of thick to thin disk mass is a decreasing
function of the circular velocity, so that more massive galaxies have
a relatively less massive thick disk. \citeauthor{Yoachim.Dalcanton.06}
(\citeyear{Yoachim.Dalcanton.06}, see caption of their
  Fig. 22) provide a simple fitting formula for this
decrease, which, although poorly constrained for massive galaxies, has
the advantage of fitting the whole mass range. Applying it for
$V_{circ}$=210 km/sec, we find a ratio of thick to thin disk masses of 0.10,
which, given the uncertainties due to the extrapolation, is in good
agreement to the value of 
0.14 we found in Sect.~\ref{subsec:disk-vs-nondisk}.  

We calculated the mean tangential velocities of G1 and G5 in an
annulus between 5 and 15 kpc from the center. This adopted radial
range is not optimum, 
since it includes mainly large radii where the difference of the two means
will be relatively small, but it was chosen so as to avoid the bar region,
where the kinematics depend mainly on the strength of the bar,
and only indirectly on the disk thickness. We find 209 and 197 km/sec
for the thin and the thick disk respectively, values which are as
expected in both sense and amplitude. They are also in agreement with the
results of \cite{Yoachim.Dalcanton.08}, who find that for the higher
mass galaxies in their sample, they fail to detect differences
between the thin and the thick disk kinematics.


\subsection{On the role of the halo gas}
\label{subsec:gas}

\begin{figure*}
  \centering
   \includegraphics[scale=0.35]{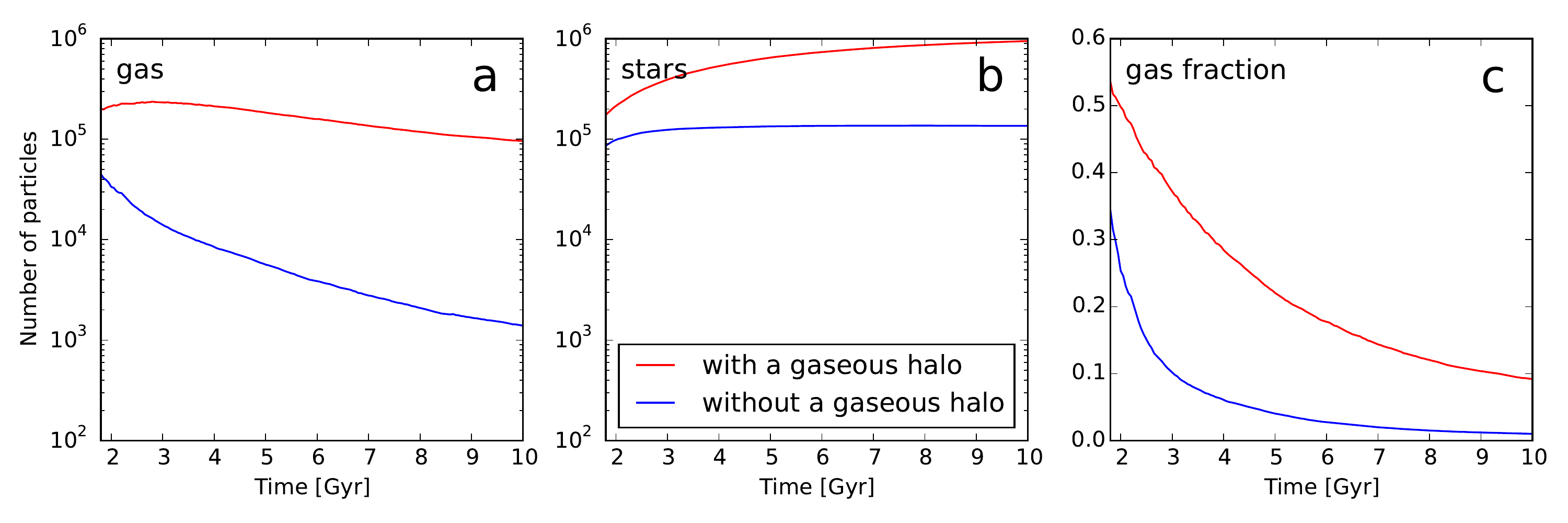}
\caption{Number of gaseous (left panel) and stellar (middle)  
  particles in the disk component as a function of time. The rightmost
  panel shows the gas fraction, again as a function of time. Red (blue) stand
  for a merging of two progenitors with (without) a gaseous
  halo (see Sect.~\ref{subsubsec:gas-origin} for a description).
}. 
\label{fig:hot-halo-effect}
\end{figure*}

\begin{figure*}
  \centering
   \includegraphics[scale=0.3]{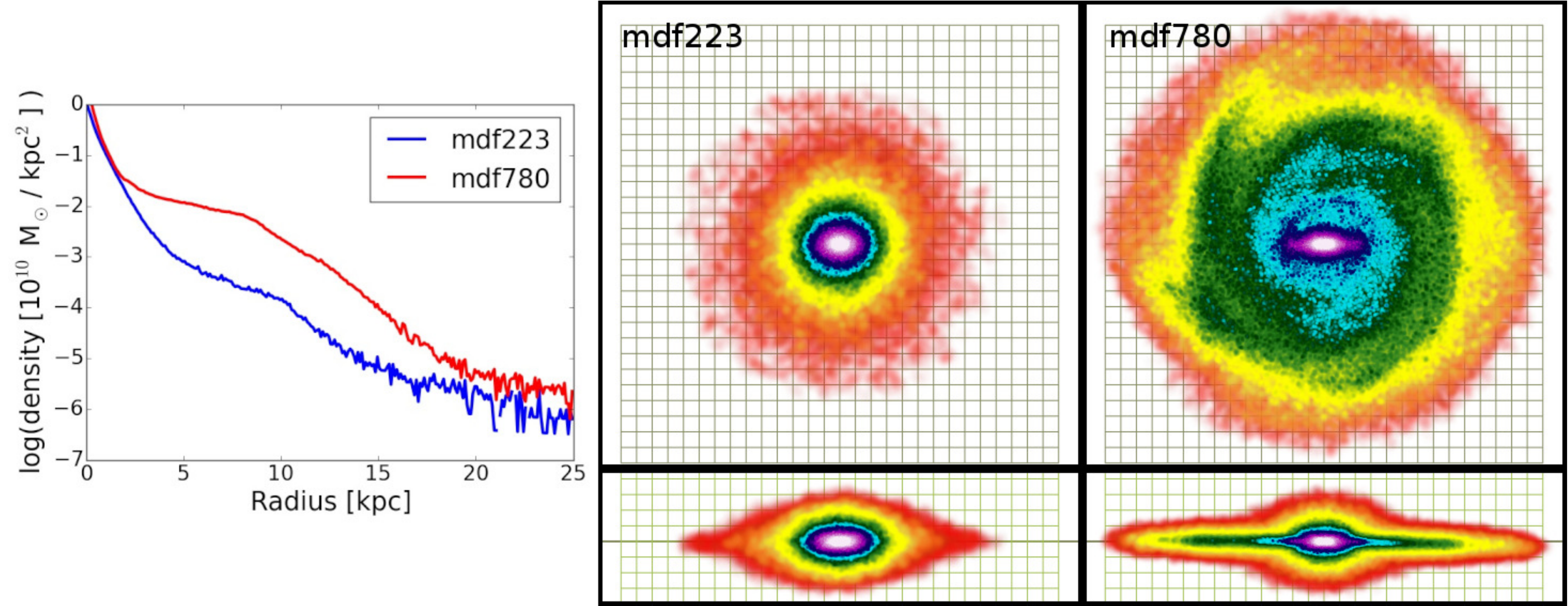}
\caption{Comparison of two simulations, one with (mdf780) and the
  other without (mdf223) a hot gaseous halo, both at time $t$=10 Gyr.
  Left: The radial projected stellar surface density 
  profiles. Middle: Face-on (upper) and edge-on (lower subpanels)
  views for mdf223. Right: Same 
  for mdf780. Note the big difference in the disk extent and in its mass
  relative to the classical bulge component.
}. 
\label{fig:mdf780mdf223}
\end{figure*}

As already discussed in Sect.~\ref{sec:sims} and \ref{subsec:early-isolated}, 
our protogalaxies acquire a baryonic
(stellar plus gaseous) disk before the merging. Thus our simulations
describe the merging of two disk protogalaxies with a halo composed of
DM and gas. We will here examine the role of the halo gas on the
formation and evolution on the ensuing disk galaxy by tracking the
origin of gaseous and stellar particles and by comparing cases
with and without halo gas. We will thus use two different approaches:

\subsubsection{Tracking the origin of gaseous and stellar particles in
  the disk}
\label{subsubsec:gas-origin}

In this first approach we will use mdf732 to disentangle 
the gas which would be present in simulations
with a gaseous disk but no gas in the halo, 
from the gas which is present 
in simulations including a gaseous halo (i.e. all the gas in our
simulations). We track each of these two separately and 
follow how each one evolved up to 10 Gyr, i.e. the end of the
run. We also take into account that each gas particle may have stayed
in gaseous  
form or it may have turned into a stellar particle.     

In order to disentangle the gas origin 
we use a snapshot at $t$=0.85 Gyr. This 
roughly corresponds to the time of the apocenter following the first 
pericenter. We also tried other times, between 0.7 and 0.9 Gyr 
and found
qualitatively the same and quantitatively very similar results. Since
the equatorial planes of the two protogalaxies coincide with the
orbital plane, we can roughly define as halo gas all gas
particles with $|\Delta z|>$1 kpc, so that the gas which would be present
in a simulation with no gaseous halo is only the gas with
$|{\Delta}z|<$1 kpc. The latter includes the gas in the two protogalactic disks
before the merging
as well as the gas in the tails formed during the interaction. 
It is important to include the gas in the tails because it
moves initially outwards but eventually  
turns back at larger distances from the center and falls back 
towards the remnant disk. We can thus roughly assume that by tracking
the gas with $|{\Delta}z|<$1 kpc we follow the gas that would be present
in a simulation with no gas in the halo. Of course the slab
$|{\Delta}z|<$1 kpc will also 
contain some halo gas, so that the above rough decomposition gives
an upper limit of what the gas in cases with no halo gas can do. This
upper limit, however, should be quite near the actual value.

Fig.~\ref{fig:hot-halo-effect}a shows the mass found in
the remnant disk still in gaseous form as a function of time, both
including (red line) and excluding (blue line) the gas in the halo. 
The former shows a much weaker decrease with time (less than a factor
of 2.5 between 3 and 10 Gyr) than the latter (more than a factor of
10 in the same time). This is
clearly due to the gas accreted from the 
halo, which leads to much more gas-rich disks all through the
simulation and a slower gas mass decrease. 

Fig.~\ref{fig:hot-halo-effect}b gives a similar comparison, but now
for the stars. Here we compare the stellar mass formed in cases with
no gaseous halo (blue line) to all disk stars (independent of the 
origin of the gas they were born from -- red line). The
latter shows a steady increase with time, compared to the former which
is nearly constant. More specifically we witness an increase by a
factor of 2.4 for the latter compared to barely a factor of 1.1
for the former. 

Fig.~\ref{fig:hot-halo-effect}c gives the ratio of gaseous to
total baryonic mass as a function of time. This decreases from
0.37 at 3 Gyr to 0.092 at 10 Gyr when halo gas accretion is included,
while it decreases from 0.1 to 0.01 
in the same time range when accretion from the halo is not included. 

To summarise, both the stellar and the 
gaseous disk are more massive when the accretion from the halo gas is
taken into account, as expected, while the gas fraction in the disk is
more compatible with that of spiral galaxies.

\subsubsection{A test simulation}
\label{subsubsec:cut-halo-gas}

To deepen our understanding of the effect of the halo gas component on
the disk galaxy formation and evolution we ran a 
further simulation. As initial conditions we used a snapshot of
simulation mdf780 at $t$=1.77 Gyr, 
in which we replaced the low density gas
particles (local density $<$ 5${\times}$10$^{-4}$ M$_\odot$ /pc$^3$)
by collisionless 
particles. These had the same mass and positions as the
gas particles they replaced, and their velocity was taken from the
velocity distribution of the halo (as e.g. in \citealt{Rodionov.AS.09}), in
order to preserve as much as possible the 
dynamical equilibrium of the system. We also verified that
this gas was essentially in the halo and that the gas in the disk and
in the tidal tails was not affected by this change.  
This is run mdf223, which we compare to mdf780 in
Fig.~\ref{fig:mdf780mdf223}. 
Even a cursory glance shows that there is
quite important differences between their mass distributions. The disk
in mdf780
is much more extended and massive
than that of mdf213. 
These results are in agreement with what we already
discussed from the analysis of mdf732 in
Sect.~\ref{subsubsec:gas-origin} and the two together show that  
the gaseous halo plays a crucial  role in the formation and
evolution of the disk.

\section{A simple scenario for the formation of disk galaxies}
\label{sec:scenario}

Putting together the results from the previous section, we can outline
the following simple scenario of disk formation via major mergers.

Two spherical protogalaxies, composed of DM and hot gas, are on an
orbit leading to a merger. From the onset of the
simulation, however, the gas in each halo cools radiatively  
and, getting out of equilibrium, falls inwards. Its density increases 
locally and the first stars form. Thus, the progenitors gradually
acquire a disk and by the time the merging starts, the two
progenitors can be described as disk protogalaxies, i.e. disk
galaxies which are smaller and more gas rich than present day galaxies.

During the merging period the potential changes drastically during a
relatively short period of time. Therefore most of the stars which were
formed before the beginning of the merging period will undergo violent
relaxation and will 
form a spheroidal bulge. Concurrently, a considerable fraction
of the gas in the progenitor falls inwards and forms stars in the
same bulge area. Most of the remaining disk gas moves outwards and forms,
together with some of the 
outermost stars, extended tails.

A number of stars, particularly those born near the end of the merging
period, will not be part of the bulge, but will form a thick disk.

After the end of this merging phase, the evolution stops being violent
and becomes secular. The material
in the tails gradually reach apocenter and fall back towards the
center of the remnant. However, most of the gas accreting on the disk
comes from the hot 
gaseous halo and forms a thin extended disk. In our three fiducial
simulations this disk is bar unstable and a bar component grows
concurrently with the disk. As in isolated disk simulations 
\citep[][for a review]{Athanassoula.15}, this bar is composed of two
parts an outer thin bar and an inner thick one. The latter is often
referred to as the boxy/peanut/ bulge but in fact is only the inner part
of the bar grown from disk instabilities. The stars forming during the
most recent 
times can be found in spirals, mainly grand design, as well as in 
a disky pseudobulge.

Very schematically, the formation of disk galaxies from a major merger
can be seen as a three-stage process. The stars born before, or at the
beginning of the merging period will undergo very strong and abrupt  
changes of the potential in which they evolve and therefore be subject to
violent
relaxation. They will constitute the classical bulge, and will be the
oldest in the galaxy. Stars born around the end of the
merging period will still 
feel considerable changes of the potential. These, however, are less
strong and abrupt than those felt by the older stars, so that they are
only strongly shuffled and end up in a thick disk component. Stars
which are born well after the end of the merging period are born in
near-circular orbits near the equatorial plane of the galaxy and form
the thin disk. Thus, according to our scenario, the sequence from
classical bulge, to thick disk, 
to thin disk should be a sequence of stellar age, with the stars in
the classical bulge being the oldest and those in the thin disk being
the youngest. The amount of perturbation they went through also
decreases along the same sequence, from violent relaxation, to simple
secular evolution. Furthermore their vertical thickness also decreases
in the same way, from the triaxial classical bulge spheroid, to a thick
and then a thin disk. 

Of course once the thin disk is even partially in place, its
evolution will be driven 
by its instabilities which will form the observed components of disk
galaxies, such as bars, spirals, lenses, rings etc.
Gas will be pushed inwards, either by the
bar or spirals, or by various asymmetries in the merging or
post-merging phases. This can form the disky pseudobulge, which will,
therefore, have a considerable amount of young stars and gas, but also
include some older stars.  

Thus the major merger scenario can account for many observational constraints.

\section{Summary and conclusions}
\label{sec:summary}

This paper is the first of a series using N-body simulations to test
whether the remnants of early major mergers can be spiral
galaxies. Our main improvement with respect to previous work on this
specific subject is that each of the progenitor galaxies in our simulations
has a hot gaseous halo component and these merge together in a hot gaseous
halo of the merger remnant. We also have a larger number of
particles, by as much as an order of magnitude compared to many previous
simulations. Furthermore, we introduced a simple but effective AGN
feedback model, which quenching star formation avoids excessive
central mass concentrations of 
the remnants, thus leading to realistic shapes of the inner part of
their rotation curves and allowing bars to form. 

Contrary to previous work on this specific subject, our initial
conditions do not mimic local galaxies, they are spherical
distributions of DM and gas. Before including them in the merging
simulations, we followed their growth in isolation and found that
during the first few Gyr they
form disk protogalaxies compatible with observations of disk galaxies
at intermediate redshifts. In particular, they are 
dynamically less relaxed and much smaller than
local galaxies, growing inside-out at a rate which is strong
initially but decreases with time. 
They are also much more gas rich than local galaxies, with a
gas fraction which decreases steadily with time, again at a faster
rate during the
earlier times. Thus what merges in our simulations are disk
protogalaxies, more compatible with observations of galaxies at
intermediate redshifts than with local ones. 

To get both qualitative and quantitative estimates of the effect of the
gaseous halo on the disk formation process in our scenario we used two
different approaches. Both showed clearly that the presence of the
gaseous halo leads to stellar and gaseous disks which are 
more massive and more extended. Star formation
continues all through the simulation, so that there will be
young stars at all times, as expected. On the contrary, if no gaseous
halo is included, the amount of gas decreases with time and star
formation grinds to a halt. Thus our improvement will allow us to
discuss elsewhere in this series the chemical evolution and population
synthesis in galactic disks. 

In our model, the progenitor disks are
destroyed by violent relaxation during the merging, but
a new disk forms in the remnant, mainly composed of material that is
gradually accreted from the halo after the merging and up
to the present time. Thus, the formation of the bulge is due to a
violent event, the merging, while the formation of the disk is
secular. Furthermore, the existence of a sizeable disk in the major merger
remnant by the end of the simulation does not argue that the disks
survived major mergers,  
but that a new disk forms after the
merging. This will have implications for the chemical composition
and colors of the galactic disk, as expected, and as we will discuss
in a future paper.

All three  of our remnant examples have 
a thin, 
extended disk and a classical bulge. The projected radial density
profiles of the disks are of type II, 
i.e. have downbending truncations. The rotation curves are flat and
show that the disk is submaximum in all three cases, 
albeit only borderline so in one of the three. The disk has substructures, which
our resolution allows us to examine. In particular we find bars,
spirals, ovals, rings and disky pseudobulges. In all cases, their
morphology is very realistic. Bars have both a thin and a thick
component, the latter being better known as a boxy/peanut
bulge. They also have ansae. Note that all three types of bulges --
classical, disky pseudobulges and boxy/peanut/X -- are
simultaneously present, as in observations \citep[e.g.][and
references therein]{Erwin.08}, as could be expected from the scenario
we present here.  
We also found ovals and realistic inner and outer rings and
pseudorings. By following the orbits of individual stellar particles
in the simulations, we found conclusive evidence that at least in one
of the three simulations the spiral is of manifold origin.

A separate thick
disk component forms naturally in our simulations, as would be
expected from the very common, if not ubiquitous appearance of these
structures in observed disk galaxies. Their properties -- such as their mass,
mean tangential 
velocities, substructures and the age of their stellar populations -- 
are compatible with both observations and previous theoretical work. 

We introduced a new method for distinguishing the bulge stellar
particles from those of the disk and its substrucures. This couples
information from stellar ages, kinematics, morphology, spatial distribution and
projected surface density profiles. It is rather precise, albeit not
straightforward. We compare the results with those of a straightforward 
method proposed by \cite{Hopkins.CYH.09} and find very good agreement for
cases with no bar. This agreement, however, becomes less good for
cases with bars and can be unsatisfactory for cases with 
very strong bars.

In our three fiducial simulations, we find that at $t$=10
Gyr, the mass of the classical bulge is, on average, between 10 and 20\% 
of the total stellar mass, i.e. values much smaller than in previous
works, as required in order to agree with the lower $B/T$ values of
spiral galaxies. This 
improvement is a corollary of the existence of a hot gaseous halo in
our 
initial conditions, as this entails a slow formation and evolution of
a massive disk. Indeed the mass of the classical bulge is set by the number
of stars formed before the merging, i.e. depends relatively little on
the existence of the gaseous halo, while the thin disk
is much more massive in cases with such a component.

This, however, does not mean that all
disk galaxies which are remnants of major
mergers of protogalaxies with a gaseous halo will have such small
classical bulges. We will present many examples with more massive
classical bulges elsewhere. Note also that we were unable to form
disk galaxies with no classical bulge at all. This is in agreement
with the discussion in \cite{Kormendy.DBC.10}, which argues that such
bulgeless galaxies would be mainly found in low density environments
where major mergers would be rather rare.

We are also able to build a sequence of components, according to
the time of formation of their stars. 
The oldest stars are found in the classical bulge, since they are
actually formed before the merging. The stars in the thick disk are
the next to form, followed by the intermediate age stars
in the disk and bar. The youngest stars can be found in the spirals
and in a disky pseudobulge. Furthermore, based on the ensemble of our
results, we are able to propose a simple scenario for the formation
of disk galaxies in major mergers.
 
We made a number of comparisons of our results with observations of
spiral galaxies and found nothing that could exclude the possibility of
forming such galaxies in major mergers of disk galaxies. We can thus consider  
the work presented here as a first ``proof of
concept'' that remnants of major mergers of two disk galaxies with a hot
gaseous halo component can be spiral galaxies. Elsewhere we will
discuss more examples with various morphologies, $B/T$ ratios, and
kinematics, as well as specific dynamical aspects. We will thus be
able to make more complete and detailed comparisons with
observations which, in turn, will allow us to answer fully whether
major mergers are a possible way of making disk galaxies, or not. 

\vspace{5 mm} 
\section*{APPENDIX}

In this appendix we briefly review the nomenclature we use in this
paper regarding bulges. More extended discussion on this and
specific references can be found in 
\cite{Kormendy.Kennicutt.04}, \citeauthor{Athanassoula.05}
(\citeyear{Athanassoula.05},\citeyear{Athanassoula.15}), 
  \cite{Drory.Fisher.07}, and \cite{Fisher.Drory.15}. 

Bulges are not a homogeneous class of objects.
We here distinguish between classical
bulges and disky pseudobulges. The former have a spheroidal shape and
a S\'ersic projected density profile with a large exponent, above say 2.5.
They rotate relatively little, with $V_{max}$/$\sigma$ values that are
consistent with isotropic oblate rotators.
On the other hand, disky pseudobulges have the shape of a thin disk
and a S\'ersic profile exponent below 2.5. They show rotation and also
they harbour structures like inner bars and inner spirals. When we
discuss here the $B/T$ ratio, we refer to the ratio of the
classical bulge mass to the total stellar mass.

Boxy/peanut/X bulges are called Boxy/peanut/X because of their shape
and are called bulges because they protrude out of the galactic
equatorial plane. Physically, however, they are just the thick part of
the bar. The same structure, seen face-on, is often called a
barlens. Although they are clearly not classical bulges, boxy/peant/X
bulges should {\it not} be referred to as pseudobulges, so as not to be
confused with disky pseudobulges. 

\vspace{5 mm} 
\section*{Acknowledgments}

We thank Albert Bosma and F. Hammer for stimulating discussions, and  
Volker Springel for 
the version of \textsc{gadget} used here. We also thank an anonymous
referee for their useful report which helped us improve the
presentation of this paper.
We acknowledge financial support from CNES
(Centre National d'Etudes Spatiales, France) and
from the EU Programme FP7/2007-2013/, under REA grant 
PITN-GA-2011-289313, as well as 
HPC resources from GENCI/TGCC/CINES 
(Grants x2013047098 and x2014047098) 
and from Mesocentre of Aix-Marseille-Universit\'e
(program  DIFOMER).


\begin{thebibliography}{99}

\expandafter\ifx\csname natexlab\endcsname\relax\def\natexlab#1{#1}\fi




\bibitem[\protect\citeauthoryear{Abadi et al.}{2003}]{Abadi.NSE.03}
    Abadi, M., Navarro, J., Steinmetz, M., Eke, V. 2003, ApJ, 597, 21  

\bibitem[\protect\citeauthoryear{{Athanassoula}}{{Athanassoula}}{1983}]{Athanassoula.83}
  {Athanassoula}, E. 1983, in Internal kinematics and dynamics of
    galaxies, ed. E. Athanassoula, Dordrecht, D. Reidel Publishing Co., 243

\bibitem[\protect\citeauthoryear{{Athanassoula}}{{Athanassoula}}{1984}]{Athanassoula.84}
  {Athanassoula}, E. 1984, PhR, 114, 319

\bibitem[\protect\citeauthoryear{{Athanassoula}}{{Athanassoula}}{1988}]{Athanassoula.88}
  {Athanassoula}, E. 1988, in Towards understanding galaxies at large
  redshift, Dordrecht, Kluwer Academic Pub., 1988, 111

\bibitem[\protect\citeauthoryear{{Athanassoula}}{{Athanassoula}}{1992}]{Athanassoula.92}
  {Athanassoula}, E. 1992, MNRAS, 259, 345

\bibitem[\protect\citeauthoryear{{Athanassoula}}{{Athanassoula}}{2003}]{Athanassoula.03} {Athanassoula}, E. 2003, MNRAS, 341, 1179 

\bibitem[\protect\citeauthoryear{Athanassoula}{2005}]{Athanassoula.05} {Athanassoula}, E. 2005, MNRAS, 358, 1477

\bibitem[\protect\citeauthoryear{{Athanassoula}}{{Athanassoula}}{2007}]{Athanassoula.07}
  {Athanassoula}, E. 2007, MNRAS, 377, 1569

\bibitem[\protect\citeauthoryear{{Athanassoula}}{{Athanassoula}}{2012}]{Athanassoula.12} {Athanassoula}, E. 2012, MNRAS, 426, 46

\bibitem[\protect\citeauthoryear{{Athanassoula}}{{Athanassoula}}{2013}]{Athanassoula.13}
  {Athanassoula}, E. 2013, in Secular Evolution of Galaxies,
  ed. J. Falc\'on-Barroso \& J. H. Knapen, Cambridge, UK: Cambridge University Press, 305

\bibitem[\protect\citeauthoryear{{Athanassoula}}{{Athanassoula}}{2015}]{Athanassoula.15}
  {Athanassoula}, E. 2015, in Galactic Bulges, ed.
  E. Laurikainen, R. Peletier and D. Gadotti, Springer Verlag,
  Germany, p. 391

\bibitem[\protect\citeauthoryear{Athanassoula et al.}{1987}]{Athanassoula.BP.87}Athanassoula, E., Bosma, A.,
  Papaioannou, S. 1987, A\&A, 179, 23

\bibitem[\protect\citeauthoryear{Athanassoula et al.}{Athanassoula et al.}{2015}]{Athanassoula.LSB.15}
  Athanassoula, E., Laurikainen, Salo, H., Bosma, A. 2015, MNRAS, 454, 3843


\bibitem[\protect\citeauthoryear{{Athanassoula} \& {Misiriotis}}{{Athanassoula} \& {Misiriotis}}{2002}]{Athanassoula.Misiriotis.02} {Athanassoula}, E., {Misiriotis}, A. 2002, MNRAS, 330, 35

\bibitem[\protect\citeauthoryear{Athanassoula et al.}
   {2009}]{Athanassoula.RGM.09} Athanassoula, E., Romero-G\'omez, M.,
  Masdemont, J. J. 2009, MNRAS, 394, 67  

\bibitem[\protect\citeauthoryear{{Aumer} \&
    {White}}{2013}]{Aumer.White.13} Aumer, M., White, S. 2013, MNRAS, 428, 1055




\bibitem[\protect\citeauthoryear{Barnes}{1998}]{Barnes.98} Barnes,
  J. 1998, in Galaxies: Interactions and Induced Star Formation,
  ed. D. Friedli, L. Martinet and D. Pfenniger, Springer-Verlag
  Berlin/Heidelberg, p. 275

\bibitem[\protect\citeauthoryear{Barnes}{2002}]{Barnes.02} Barnes,
  J. 2002, MNRAS, 333, 481

\bibitem[\protect\citeauthoryear{Bershady et
    al.}{2011}]{Bershady.MVWAS.11} Bershady, M., Martinsson, T.,
  Verheijen, M., et al. 2011, ApJL,
  739, 47

\bibitem[\protect\citeauthoryear{Bondi}{1952}] {Bondi.52}
Bondi, H. 1952, MNRAS, 112, 195

\bibitem[\protect\citeauthoryear{Bondi \& Hoyle}{1944}]{Bondi.Hoyle.44} 
Bondi, H., Hoyle, F. 1944, MNRAS, 112, 195

\bibitem[\protect\citeauthoryear{Borlaff et al.}{2014}]{Borlaff.EMRPQTPGZGB.14} 
	Borlaff, A., Eliche-Moral, M. C., Rodr\'iguez-P\'erez, C., et al. 2014, A\&A, 570, 103

\bibitem[\protect\citeauthoryear{Bouwens et al.}{2004}]{Bouwens.IBBF.04}
Bouwens, R., Illingworth, G., Blakeslee, J. 2004, ApJ, 611, L1


\bibitem[\protect\citeauthoryear{Brook et al.}{2012}] {Brook.SGRWQ.12}
Brook, C., Stinson, G., Gibson, et al., 2012 MNRAS, 419, 771

\bibitem[\protect\citeauthoryear{Brooks \&
    Christensen}{2015}] {Brooks.Christensen.15} Brooks, A.,
  Christensen, C. 2015, in Galactic Bulges, ed.
  E. Laurikainen, R. Peletier and D. Gadotti, Springer Verlag,
  Germany, 317

\bibitem[\protect\citeauthoryear{Buta}{1995}] {Buta.95}
Buta, R. 1995, ApJS, 96, 39

\bibitem[\protect\citeauthoryear{Buta et al.}{2015}] {Buta.P.15}
Buta, R., Sheth, K., Athanassoula, E., et al. 2015, ApJS, 217, 32


\bibitem[\protect\citeauthoryear{Casertano \& Hut}{1985}]
  {Casertano.Hut.85} 
{Casertano, S., Hut, P. 1985, ApJ, 298, 80}

\bibitem[\protect\citeauthoryear{Colin et al.}{2006}]
  {Colin.VK.06} 
{Colin, P., Valenzuela, O., Klypin, A. 2006, ApJ, 644, 687}

\bibitem[\protect\citeauthoryear{Comer\`on et al.}{2011b}] {Comeron.thick-global-1.11}
Comer\`on, S., Elmegreen, B. G., Knapen, J. H., et al. 2011b, ApJ, 741, 28

\bibitem[\protect\citeauthoryear{Comer\`on et al.}{2012}] {Comeron.thick-global-2.12}
Comer\`on, S., Elmegreen, B. G., Salo, H., et al. 2012, ApJ, 759, 98

\bibitem[\protect\citeauthoryear{Comer\`on et al.}{2011a}] {Comeron.thick-4244.11}
Comer\`on, S., Knapen, J. H., Sheth, K., et al. 2011a, ApJ, 729, 18
  
\bibitem[\protect\citeauthoryear{Comer\`on et al.}{2015}]{Comeron.SJLY.15}
Comer\'on, S., Salo, H., Janz, J., Laurikainen, E., Yoachim, P. 2015, A\&A, 584, 34

\bibitem[\protect\citeauthoryear{Conselice et
    al.}{2012}]{Conselice.MBG.12}{Conselice C., Mortlock A., Bluck
  A. F., Gr\"utzbauch R., Duncan, K. 2013, MNRAS, 430, 1051} 

\bibitem[\protect\citeauthoryear{Cox et al.}{2006}]{Cox.JPS.06} Cox,
  T. J., Jonsson, P., Primack, J. R., Somerville, R. S. 2006, MNRAS,
  373, 1013

\bibitem[\protect\citeauthoryear{Cuddeford}{1991}] {Cuddeford.91}
Cuddeford, P. 1991, MNRAS, 253, 414

\bibitem[\protect\citeauthoryear{Cullen \& Dehnen}{2010}]
{Cullen.Dehnen.10} Cullen, L., Dehnen, W. 2010, MNRAS, 408, 669



\bibitem[\protect\citeauthoryear{Daddi et al.}{2010}]{Daddi.BWDCDEMROSKS.10}
  Daddi, E., Bournaud, F., Walter, F., et al. 2010, ApJ, 713, 686

\bibitem[\protect\citeauthoryear{Dahlen et al.}{2007}] {Dahlen.MDFGKR.07}
Dahlen, T., Mobasher, B., Dickinson, M. et al. 2007, ApJ, 654, 172

\bibitem[\protect\citeauthoryear{Diaz et al.}{2015}] {Diaz.GSLHE.15}
D\'iaz-García, S., Salo, H., Laurikainen, E., Herrera-Endoqui, M.
2015, arXiv:1509.06743

\bibitem[\protect\citeauthoryear{Dobbs \& Baba}{2014}] {Dobbs.Baba.14}
Dobbs, C., Baba, J. 2014, PASA, 31, id.e035 

\bibitem[\protect\citeauthoryear{Drory \& Fisher}{2007}]{Drory.Fisher.07} 
{Drory}, N., {Fisher}, D.~B.\ 2007, ApJ, 664, 640 


\bibitem[\protect\citeauthoryear{Elmegreen et al.}{2005}]
  {Elmegreen.EVFF.05} Elmegreen, B. G., Elmegreen, D. M., Vollbach,
  D. R.,  Foster, E. R., Ferguson, T. E. 2005, ApJ, 634, 101

\bibitem[\protect\citeauthoryear{Erb et al.}{2006}]{Erb.SSPRA.06}
  {Erb} D.~K., {Steidel} C.~C., {Shapley} A.~E., et al. 2006, \apj, 646, 107 

\bibitem[\protect\citeauthoryear{Erwin}{2008}]{Erwin.08} Erwin,
  P. 2008, in Formation and Evolution of Galaxy Bulges, ed. M.
  Bureau, E. Athanassoula, B. Barbuy, IAU Symp., 245, 113

\bibitem[\protect\citeauthoryear{Erwin et al.}{2005}]{Erwin.BP.05} Erwin, P., Beckman, J. E., Pohlen, M. 2005, ApJ, 626, L81

\bibitem[\protect\citeauthoryear{Eskridge et al.}{2000}]{Eskridge.P.00} 
Eskridge, P.~B., Frogel, J.~A., Pogge, R.~W., et al.\ 2000, AJ, 119,
536 


\bibitem[\protect\citeauthoryear{Ferrarese \& Merritt}{2000}]{Ferrarese.Merritt.00} Ferrarese L., Merritt D. 2000, ApJ, 539, L9

\bibitem[\protect\citeauthoryear{Ferguson et al.}{2004}]
  {Ferguson.P.04} Ferguson, H.,
  Dickinson, M., Giavalisco, M. 2004, ApJ, 600, L107

\bibitem[\protect\citeauthoryear{Fisher \& Drory}{2015}]{Fisher.Drory.15}
  Fisher, D., Drory, N. 2015, in Galactic Bulges, ed.
  E. Laurikainen, R. Peletier and D. Gadotti, Springer Verlag,
  Germany, 41 

\bibitem[\protect\citeauthoryear{Freeman}{1970}]{Freeman.70} Freeman,
  K. C. 1970, ApJ, 160, 811


\bibitem[\protect\citeauthoryear{Gebhardt et
    al.}{2000}]{Gebhardt.p.00} Gebhardt K., Bender, R., Bower, G., et al. 2000, ApJ, 539, L13

\bibitem[\protect\citeauthoryear{Genzel et al.}{2015}]{Genzel.p.15}
  Genzel, R.~A., Tacconi, L. J., Lutz, D., et al. 2015, ApJ, 800, 20

\bibitem[\protect\citeauthoryear{Governato et
    al.}{2009}]{Governato.BBMWJSPCWQ.09} Governato, F., Brook, C. B.,
  Brooks, A. M., et al. 2009, MNRAS, 398, 312





\bibitem[\protect\citeauthoryear{Hammer et
    al.}{2005}]{Hammer.FEZLC.05} Hammer, F., Flores, H., Zheng, X. Z.,
  Liang, Y. C., Cesarsky, C. 2005, A\&A, 430, 115

\bibitem[\protect\citeauthoryear{Hammer et
    al.}{2009a}]{Hammer.FPYARD.09} Hammer, F., Flores, H., Puech, M.,
  et al. 2009a, A\&A, 507, 1313

\bibitem[\protect\citeauthoryear{Hammer et al.}{2009b}]{Hammer.FYAPRP.09} Hammer, F., Flores, H., Yang,
  Y. B., et al. 2009b, A\&A, 496, 381

\bibitem[\protect\citeauthoryear{Hayward et al.}{2014}] {Hayward.TSHV.14}
Hayward, C., Torrey, P., Springel, V., Hernquist, L., Vogelsberger,
M. 2014, MNRAS, 442, 1992

\bibitem[\protect\citeauthoryear{Hopkins}{2013}]{Hopkins.13} Hopkins,
  P. F. 2013, MNRAS, 428, 2840

\bibitem[\protect\citeauthoryear{Hopkins}{2014}]{Hopkins.14} Hopkins,
  P. F. 2014, Astrophysics Source Code Library, record ascl:1410.003

\bibitem[\protect\citeauthoryear{Hopkins}{2015}]{Hopkins.15} Hopkins,
  P. F. 2015, MNRAS, 450, 53

\bibitem[\protect\citeauthoryear{Hopkins et al.}{2009}]{Hopkins.CYH.09} Hopkins, P. F., Cox, T. J., Younger,
  J. D., Hernquist, L. 2009, ApJ, 691, 1168

\bibitem[\protect\citeauthoryear{Hopkins et al.}{2013}]{Hopkins.CHNHM.13} 
	Hopkins, P. F., Cox, T. J., Hernquist, L., et al. 2013, MNRAS, 430, 1901

\bibitem[\protect\citeauthoryear{Hopkins \& Quataert}{2012}]{Hopkins.Quataert.10}
Hopkins, P. F., Quataert, E. 2010, MNRAS, 407, 1529

\bibitem[\protect\citeauthoryear{Hoyle \& Lyttleton}{1939}]{Hoyle.Lyttleton.39}
Hoyle, F., Lyttleton, R. A. 1939, PCPS, 35, 405







\bibitem[\protect\citeauthoryear{Kannan et al.}{2015}]{Kannan.MFMKS.15}
Kannan, R., Macci\`o, A. V., Fontanot, F., et al. 2015, MNRAS, 452, 4347

\bibitem[\protect\citeauthoryear{{Kormendy}}{{Kormendy}}{2013}]{Kormendy.13}Kormendy, J. 2013, in Secular Evolution of Galaxies, ed. Jes\'us Falc\'on-Barroso \& Johan H. Knapen, Cambridge, UK: Cambridge University Press, 2013, 1

\bibitem[\protect\citeauthoryear{Kormendy et
    al.}{2010}]{Kormendy.DBC.10} Kormendy, J., Drory, N., Bender, R.,
  Cornell, M. 2010, ApJ, 723, 54

\bibitem[\protect\citeauthoryear{Kormendy \& Kennicutt}{2004}]{Kormendy.Kennicutt.04} Kormendy J., Kennicutt R.~C., Jr. 2004, ARA\&A, 42, 603 

\bibitem[\protect\citeauthoryear{Knapen et al.}{2000}]{Knapen.SP.00} 
Knapen, J.~H., Shlosman, I., Peletier, R.~F.\ 2000, ApJ, 529, 93 


\bibitem[\protect\citeauthoryear{Laine et al.}{2014}]{Laine.plus.14}
Laine, J., Laurikainen, E., Salo, H., et al. 2014, MNRAS, 441, 1992

\bibitem[\protect\citeauthoryear{Laurikainen et al.}{2014}] {Laurikainen.SABHE.14} Laurikainen, E., Salo, H., Athanassoula, E., Bosma, A., Herrera Endoqui, M. 2014, MNRAS, 444, L80

\bibitem[\protect\citeauthoryear{Lin}{1967}]{Lin.67}
Lin, C. C. 1967, Annual Review of Astronomy and Astrophysics, 5, 453 

\bibitem[\protect\citeauthoryear{Leroy et al.}{2008}]{Leroy.WBBBMT.08}
  {Leroy} A.~K., {Walter} F., {Brinks} E., et al. 2008, \aj, 136, 2782 


\bibitem[\protect\citeauthoryear{Lotz et al.}{2010b}]{Lotz.JCP.10b}
Lotz, J., Jonsson, P., Cox, T., Primack, J. 2010, MNRAS, 404, 590

\bibitem[\protect\citeauthoryear{Lynden-Bell}{1967}]{Lynden-Bell.67}	
	Lynden-Bell, D. 1967, MNRAS, 136, 101


\bibitem[\protect\citeauthoryear{McMillan \& Dehnen}{2007}] {McMillan.Dehnen.07}
McMillan, P., Dehnen, W. 2007, MNRAS, 378, 541

\bibitem[\protect\citeauthoryear{Men{\'e}ndez-Delmestre et al.}{2007}]{Menendez.DSSJS.07} 
Men{\'e}ndez-Delmestre, K., Sheth, K., Schinnerer, E., Jarrett,
T., Scoville, N.\ 2007, ApJ, 657, 790 

\bibitem[\protect\citeauthoryear{Miller \& Bregman}{2015}]{Miller.Bregman.15}
Miller, M., Bregman, J. 2015, ApJ, 800, 14

\bibitem[\protect\citeauthoryear{Moster et al.}{2011}]{Moster.MSNC.11}
Moster, B. P., Macci\`o, A. V., Somerville, R. S., Naab, T., Cox,
T. J. 2011, MNRAS, 415, 3750

\bibitem[\protect\citeauthoryear{Mould}{2005}]{Mould.05}
Mould, J. 2005, ApJ, 129, 698

\bibitem[\protect\citeauthoryear{Mu\~noz-Mateos et
    al.}{2013}]{Munoz.plus.13} Mu\~noz-Mateos, J. C., Sheth, K., Gil
  de Paz, A., et al. 2013, ApJ, 771, 59


\bibitem[\protect\citeauthoryear{Ness et al.}{Ness et
    al.}{2013a}]{Ness.p.13a} Ness, M., Freeman, K., Athanassoula, E., et al. 2013a, MNRAS, 430, 836

\bibitem[\protect\citeauthoryear{Ness et al.}{Ness et
    al.}{2013b}]{Ness.p.13b} Ness, M., Freeman, K., Athanassoula, E., et al. 2013b, MNRAS, 432, 2092



\bibitem[\protect\citeauthoryear{Pohlen \& Trujillo}{2006}]{Pohlen.Trujillo.06} Pohlen, M., Trujillo, I. 2006, A\&A, 454, 759


\bibitem[\protect\citeauthoryear{Querejeta et al.}{2015}]{Querejeta.EMTBRPZG.15} Querejeta, M., Eliche-Moral,
  M. C., Tapia, T., et al. 2015, A\&A, 573, A78



\bibitem[\protect\citeauthoryear{Robertson et
    al.}{2006}]{Robertson.BCDMHSY.06a} Robertson, B., Bullock, J. S.,
  Cox, T. J., et al. 2006, ApJ, 645, 986

\bibitem[\protect\citeauthoryear{Rodionov et al.}{2009}] {Rodionov.AS.09}
Rodionov, S., Athanassoula, E., Sotnikova, N. 2009, MNRAS, 392, 904

\bibitem[\protect\citeauthoryear{Rodrigues et al.}{2012}]{Rodrigues.PHRF.12} Rodrigues, M., Puech, M., Hammer, F., Rothberg, B., Flores, H. 2012, MNRAS, 421, 2888

\bibitem[\protect\citeauthoryear{Romero-G\'omez et
    al.}{2006}]{Romero.GMAGG.06} Romero-G\'omez, M., Masdemont, J. J.,
Athanassoula, E., Garc\'ia-G\'omez, C. 2006, A\&A, 453, 39 


\bibitem[\protect\citeauthoryear{Sackett}{1997}]{Sackett.97}
Sackett, P. 1997, ApJ, 483, 103 

\bibitem[\protect\citeauthoryear{Sandage}{1961}]{Sandage.61} Sandage, A. 1961, The Hubble Atlas of Galaxies (Washington:Carnegie Institution)


\bibitem[\protect\citeauthoryear{Saha et al.}{2012}]{Saha.MVG.12}Saha, K.,
  Martinez-Valpuesta, I., Gerhard, O. 2012, MNRAS, 421, 333

\bibitem[\protect\citeauthoryear{Schweizer}{1998}]{Schweizer.98} Schweizer,
  F. 1998, in Galaxies: Interactions and Induced Star Formation,
  ed. D. Friedli, L. Martinet and D. Pfenniger, Springer-Verlag
  Berlin/Heidelberg, p. 105

\bibitem[\protect\citeauthoryear{Scoville et
    al.}{1991}]{Scoville.SSS.91}
Scoville, N., Sargent, A., Sanders, D., Soifer, B. 1991, ApJ, 366, 5 

\bibitem[\protect\citeauthoryear{Shen et al.}{2010}]{Shen.RKHDK.10}
Shen, J., Rich, M., Kormendy, J., et al. 2010, \apj, 720, 72

\bibitem[\protect\citeauthoryear{Silk \& Mamon}{2012}] {Silk.Mamon.12}
Silk, J., Mamon, G. 2012, RAA, 12, 917

\bibitem[\protect\citeauthoryear{Springel}{2005}]{Springel.05} Springel, V. 2005, \mnras, 364, 1105 

\bibitem[\protect\citeauthoryear{Springel \& Hernquist}{2002}]{Springel.Hernquist.02} Springel, V., Hernquist, L. 2002, \mnras, 333, 649

\bibitem[\protect\citeauthoryear{Springel \& Hernquist}{2003}]{Springel.Hernquist.03} Springel, V., Hernquist, L. 2003, MNRAS, 339, 289

\bibitem[\protect\citeauthoryear{Springel \& Hernquist}{2005}]{Springel.Hernquist.05} Springel, V., Hernquist, L. 2005, ApJ, 622, 9

\bibitem[\protect\citeauthoryear{Springel et al.}{2005}]{Springel.DMH.05} Springel, V., Di Matteo, T., Hernquist, L. 2005, MNRAS, 361, 776

\bibitem[\protect\citeauthoryear{Streich et al.}{2015}]{Streich.P.15}
Streich, D., de Jong, R., Bailin, J., et al. 2016, A\&A 585, A97
 


\bibitem[\protect\citeauthoryear{Tacconi et al.}{Tacconi et
    al.}{2010}]{Tacconi.P.10} Tacconi L., Genzel, R., Neri, R., et al. 2010, Natur, 463, 781

\bibitem[\protect\citeauthoryear{Thacker et al.}{2014}]{Thacker.MWH.14} 
Thacker, R., MacMackin, C., Wurster, J., Hobbs, A. 2014, MNRAS, 443, 1125

\bibitem[\protect\citeauthoryear{Toomre}{1981}]{Toomre.81} Toomre,
A. 1981, in The structure and evolution of normal galaxies,
Fall, S. M. and Lynden-Bell, D., ed., Cambridge and New York,
Cambridge University Press, 111

\bibitem[\protect\citeauthoryear{Toomre \&
    Toomre}{1972}]{Toomre.Toomre.72} Toomre, A., Toomre, J. 1972, ApJ,
  178, 623

\bibitem[\protect\citeauthoryear{Torrey et al.}{2012}]{Torrey.VSSH.12}
Torrey, P., Vogelsberger, M., Sijacki, D., Springel, V., Hernquist,
L. 2012, MNRAS, 427, 2224




\bibitem[\protect\citeauthoryear{Wang et al.}{2012}] {Wang.HAPYF.12}
Wang, J., Hammer, F., Athanassoula, E. et al. 2012, A\&A, 538, 121


\bibitem[\protect\citeauthoryear{Yoachim \& Dalcanton}{2006}]
  {Yoachim.Dalcanton.06} Yoachim, P., Dalcanton, J. 2006, AJ, 131, 226

\bibitem[\protect\citeauthoryear{Yoachim \& Dalcanton}{2008}]
  {Yoachim.Dalcanton.08} Yoachim, P., Dalcanton, J. 2008, ApJ, 682, 1004




\end{thebibliography}

\label{lastpage}

\end{document}